\documentclass[aoas]{imsart}
\usepackage{amsthm}
\usepackage{amsmath}
\usepackage{amssymb}
\usepackage{graphicx}
\usepackage{amsfonts}
\usepackage{verbatim}
\usepackage{colonequals}
\usepackage[authoryear]{natbib}
\usepackage[colorlinks]{hyperref}
\usepackage{color}
\usepackage{colortbl}
\usepackage{bm}
\usepackage{booktabs}
\usepackage{algorithm2e}

\usepackage{xr}
\makeatletter
\newcommand*{\addFileDependency}[1]{
  \typeout{(#1)}
  \@addtofilelist{#1}
  \IfFileExists{#1}{}{\typeout{No file #1.}}
}
\makeatother
\newcommand*{\myexternaldocument}[1]{
    \externaldocument{#1}
    \addFileDependency{#1.tex}
    \addFileDependency{#1.aux}
}

\myexternaldocument{supplementary_materials}

\startlocaldefs

\definecolor{Gray}{gray}{0.9}

\theoremstyle{plain}

\theoremstyle{remark}

\newcommand{\bs}[1]{\boldsymbol{#1}}

\def\R{\bs{R}}
\def\V{\bs{V}}
\def\U{\bs{U}}

\def\u{\bs{u}}

\def\I{\bs{I}}

\def\x{\bs{x}}

\def\w{\bs{w}}
\def\a{\bs{a}}

\def\e{\bs{e}}

\def\S{\bs{S}}

\def\btheta{\bs{\theta}}

\def\zero{\bs{0}}
\def\one{\bs{1}}
\def\btheta{\bs{\theta}}
\def\bpi{\bs{\pi}}
\def\s{\bs{s}}
\def\P{P}

\def\cS{\mathcal{S}}

\def\cU{\mathcal{U}}

\def\tr{\text{tr}}

\DeclareMathOperator*{\argmax}{argmax}
\DeclareMathOperator*{\argmin}{argmin}

\endlocaldefs

\begin{document}

\begin{frontmatter}
\title{Improved methods for empirical Bayes multivariate multiple
    testing and effect size estimation}
\runtitle{Improved empirical Bayes multivariate multiple testing}
\runauthor{Y. Yang et al}

\begin{aug}

\author[A]{\fnms{Yunqi}~\snm{Yang}},
\author[B]{\fnms{Peter}~\snm{Carbonetto}}
\author[C]{\fnms{David}~\snm{Gerard}}
\and
\author[B,D]{\fnms{Matthew}~\snm{Stephens}\ead[label=e1]{mstephens@uchicago.edu}}
\address[A]{Committee on Genetics, Genomics and System Biology,
University of Chicago}

\address[B]{Department of Human Genetics,
University of Chicago}

\address[C]{Department of Mathematics and Statistics, American University}

\address[D]{Department of Statistics,
University of Chicago\printead[presep={,\ }]{e1}}

\end{aug}

\begin{abstract}
Estimating the sharing of genetic effects across different conditions
is important to many statistical analyses of genomic data.
The patterns of sharing arising from these data are often highly
heterogeneous.
To flexibly model these heterogeneous sharing patterns,
\cite{urbut2019flexible} proposed the multivariate adaptive shrinkage
(MASH) method to jointly analyze genetic effects across multiple
conditions. However, multivariate analyses using MASH (as well as
other multivariate analyses) require good estimates of the sharing
patterns, and estimating these patterns efficiently and accurately
remains challenging. Here we describe new empirical Bayes methods that
provide improvements in speed and accuracy over existing methods. The
two key ideas are: (1) adaptive regularization to improve accuracy in
settings with many conditions; (2) improving the speed of the model
fitting algorithms by exploiting analytical results on covariance
estimation. In simulations, we show that the new methods provide
better model fits, better out-of-sample performance, and improved
power and accuracy in detecting the true underlying signals. In an
analysis of eQTLs in 49 human tissues,
our new analysis pipeline achieves better model fits and better
out-of-sample performance than the existing MASH analysis pipeline.
We have implemented the new methods, which we call ``Ultimate
Deconvolution'', in an R package, udr, available on GitHub.
\end{abstract}


\end{frontmatter}

\section{Introduction}
\label{sec1}


The problem of testing and estimating effect sizes for many units in
multiple conditions (or on multiple outcomes) arises frequently in
genomics applications.  Examples include assessing the effects of many
expression quantitative trait loci (eQTLs) in multiple tissues
\citep{gtex2015genotype} and assessing the effects of many genetic
variants on multiple traits \citep{zhou2014efficient,
  pickrell2016detection, turchin2019bayesian, udler2018type,
  mvsusie}. The simplest approach to assessing effects in multiple
conditions is to analyze each condition separately. However, this
fails to exploit sharing or similarity of effects among
conditions. For example, a genetic variant that increases expression
of a particular gene in the heart may similarly increase expression in
other tissues, particularly in tissues that are related to heart. Such
sharing of effects could be exploited to improve power and estimation
accuracy by borrowing information across conditions. This is, in
essence, the motivation for meta-analysis methods
\citep{willer2010metal, han2011random, wen2014bayesian}, and more
generally it motivates consideration of multivariate approaches to
multiple testing and effect size estimation \citep{urbut2019flexible}.

While borrowing information across conditions may be a natural idea,
getting it to work well in practice requires confronting some thorny
issues. One challenge is that the extent to which effects are shared
among conditions will vary among data sets; for example, data sets
involving very similar conditions may have very high levels of
sharing, whereas data sets involving very different conditions may
exhibit little to no sharing. Furthermore, some data sets may include
some conditions that are very similar and others that are very
dissimilar, and these differences may be difficult to specify in
advance. Assessing the sharing and similarity of effects
among conditions may also be an important goal in itself.

Empirical Bayes (EB) approaches (e.g., \citealt{flutre2013statistical,
  urbut2019flexible}) provide an attractive way to confront these
challenges. EB methods estimate a prior distribution that captures the
sharing or similarity of effects among the conditions, then, using
Bayes theorem, they combine the prior with the observed data to
improve the effect estimates. The methods proposed in
\cite{urbut2019flexible} and implemented in the R package mashr assume
a mixture of multivariate normal distributions for the prior, which
has the twin advantages of being both both flexible and
computationally convenient. Indeed, mashr has been used to analyze
very large data sets involving many conditions
\citep{mvsusie, lin2024genetic, urbut2021bayesian,barbeira2020fine,
  araujo2023multivariate, li2022rna, soliai2021multi, natri2024cell,
  bonder2021identification}.

Despite this, these methods still have considerable limitations; in
particular, fitting a mixture of multivariate normals prior with
unknown covariance matrices raises statistical and computational
challenges. \cite{urbut2019flexible} used a two-stage procedure to
deal with (or sidestep) these challenges: in the first stage, the
covariance matrices in the prior were estimated by maximum-likelihood
on a subset of the data (using the ``Extreme Deconvolution'' algorithm
of \citealt{bovy2011extreme}); next, given the covariances estimated
in the first stage, the second stage estimated the mixture proportions
in the prior by maximzing the likelihood from all the data (using the
fast optimization algorithms of \citealt{kim2020fast}). The second
stage is a convex optimization problem, and can be solved efficiently
and reliably for very large data sets. The first stage, however,
presents several challenges, including that the Extreme Deconvolution
(ED) algorithm can be slow to converge, the results of running ED are
often sensitive to initialization, and the estimated covariance
matrices can be quite unstable, particularly when the number of
conditions, $R$, is large relative to sample size, $n$. These
challenges motivated this work, which is a closer examination of these
challenges from both a statistical perspective and a computational
one. One of the contributions of this study is a new algorithm, which
we call ``truncated eigenvalue decomposition'' (TED). TED often
converges much faster than ED. (Noting that ED applies to some
settings where TED does not.) We also explore the use of simple
regularization schemes that can improve accuracy compared with
maximum-likelihood estimation, particularly when the sample size $n$
is small and the number of conditions $R$ is large (that is, the ratio
$R/n$ is large). And we highlight some problems that arise from using
low-rank covariance matrices, which was a strategy previously
suggested in \cite{urbut2019flexible} to reduce the number of
estimated parameters.  We provide an R package, udr (``Ultimate
Deconvolution in R''), available at
\url{https://github.com/stephenslab/udr}, which implements all these
methods described here within a convenient, user-friendly interface,
and that interacts well with our previous R package, mashr
\citep{urbut2019flexible}. The version of the R package used to
perform the analyses in this manuscript is also available for download
as a supplementary file \citep{supplementary_zip}.

The paper is structured as follows. Section \ref{sec:notation}
summarizes notation used in the mathematical expressions throughout
the paper. Section \ref{sec:model} formally introduces the models,
priors and regularization schemes considered, and Section
\ref{sec:fitting} describes the model-fitting algorithms and
procedures.
Section \ref{sec:simulation} evaluates the performance of the different
methods and algorithms on data sets simulated under a variety of
scenarios. Section \ref{sec:application} applies the improved methods
to an analysis of a large, multi-tissue eQTL data set from the GTEx
Project. Finally, Section \ref{sec:discussion} discusses the promise
and limitations of our methods, and future directions.

\section{Notation used in the mathematical expressions}
\label{sec:notation}

For the mathematical expressions below, we use bold, capital letters
($\bs{A}$) to denote matrices; bold, lowercase letters ($\bs{a}$) to
denote column vectors; and plain, lowercase letters ($a$) to denote
scalars. For a matrix ${\bm A}$, $\|\bs{A}\|_{\ast}$ denotes its
nuclear norm, $\bs{A}^T$ denotes its transpose, $\bs{A}^{-1}$ denotes
its inverse, $\bs{A}^{-T}$ denotes the inverse of $\bs{A}^T$,
$\tr(\bs{A})$ denotes the trace, and $|\bs{A}|$ denotes the matrix
determinant.
%
%
We use $\zero$ and $\one$ to denote the vectors whose elements are all
zeros and all ones respectively; $\e_r$ for the standard basis vector
$\e_r = (0, \dots, 0, 1, 0, \dots, 0)$ with the 1 appearing in the
$r$th position; $\I_R$ is the $R \times R$ identity matrix; and
$\mbox{diag}(\a)$ denotes the diagonal matrix in which the diagonal
entries are given by the entries of vector $\a$. We use $\mathbb{R}$
for the set of real numbers, $\mathbb{R}^{R}$ for the set of
real-valued vectors of length $R$, and $\mathbb{R}^{n \times m}$ for
the set of real-valued $n \times m$ matrices.  We use $N(\mu,
\sigma^2)$ to denote the normal distribution on $\mathbb{R}$ with mean
$\mu$ and variance $\sigma^2$, and $N(\,\cdot\,; \mu, \sigma^2)$
denotes its density.  And we use $N_R(\bm{\mu}, \bm{\Sigma})$ for the
multivariate normal distribution on $\mathbb{R}^R$ with mean $\bm{\mu}
\in \mathbb{R}^R$ and $R \times R$ covariance matrix $\bm{\Sigma}$,
and $N_R(\,\cdot\,; \bm{\mu}, \bm{\Sigma})$ denotes its density.  We
use $\P^{+}_R \subseteq \mathbb{R}^{R \times R}$ to denote the set of
real-valued $R \times R$ (symmetric) positive semi-definite
matrices. $\cS_R \subseteq \mathbb{R}^R$ denotes the $R$-dimensional
simplex.

\section{The empirical Bayes multivariate normal means model}
\label{sec:model}

In this section, we define the ``empirical Bayes multivariate normal
means'' (EBMNM) model. We describe several variations of this model
that involve constraints or penalties on the model parameters.

The EBMNM model assumes that we observe vectors $\x_j \in
\mathbb{R}^R$, that are independent, noisy, normally-distributed
measurements of underlying true values $\btheta_j \in \mathbb{R}^R$:
\begin{equation}
\label{eq:mvnm}
\x_j \mid \btheta_j \sim N_R(\btheta_j,\V_j),
\quad j = 1, \ldots, n,
\end{equation}
in which the covariances $\V_j \in \P^{+}_R$ are assumed to be known
and invertible.
Our ultimate goal is to perform inference for the unknown means
$\btheta_j$ from the observed data $\x_j$. This model is a
natural generalization of the well-studied (univariate) normal means
model \citep{robbins1951, efron1972limiting, stephens2017false,
  bhadra2019lasso, Johnstone, lei-thesis}, and so we call it the
``multivariate normal means model''.
%
%
An important special case is when the measurement error distribution
is the same for all observations; i.e., $\V_j = \V$, $j = 1, \ldots,
n$. We refer to this as the ``homoskedastic'' case.
Some of our computational methods are designed specifically for the
homoskedastic case, which is easier to solve than the
``heteroskedastic'' case.
%
%


The EBMNM model \eqref{eq:mvnm} further assumes that the unknown means
are independent and identically distributed draws from some
distribution, which we refer to as the ``prior distribution''. While
other choices are possible, here we assume the prior distribution is a
mixture of zero-mean multivariate normals:
\begin{equation}
\label{eq:prior}
p(\btheta_j \mid \bpi,\cU) =
\sum_{k=1}^K \pi_k N_R(\btheta_j; \zero, \U_k),
\end{equation}
where $\bpi \in \cS_K$ is the set of mixture proportions, $\U_k \in
\P_R^{+}$ are covariance matrices, and $\cU \colonequals \{\U_1,
\dots, \U_K\}$ denotes the full collection of covariance matrices.


Combining \eqref{eq:mvnm} and \eqref{eq:prior} yields the marginal
distribution
\begin{equation}
\label{eq:marginal}
p(\x_j \mid \bpi,\cU) =
\sum_{k=1}^K \pi_k N_R(\x_j; \zero, \U_k + \V_j),
\end{equation}
and the marginal log-likelihood,
\begin{align}
\ell(\bpi,\cU) \colonequals&\;
\sum_{j=1}^n \log p(\x_j \mid \bpi,\cU) \nonumber \\
=&\; \sum_{j=1}^n \log \sum_{k=1}^K \pi_k N_R(\x_j; \zero, \U_k + \V_j)
\end{align}

The EB approach to fitting the model (\ref{eq:mvnm}--\ref{eq:prior})
proceeds in two stages:
\begin{enumerate}
\setlength{\itemsep}{6pt}  

\item Estimate the prior parameters $\bpi, \cU$ by maximizing a
penalized log-likelihood,
\begin{equation}
\label{eq:penloglik}
(\hat{\bpi}, \hat{\cU}) \colonequals
\argmax_{\bpi \,\in\, \cS_K,\, \U_k \,\in\, \P^{+,k}_R}
\ell(\bpi, \cU) - \sum_{k=1}^K \tilde{\rho}(\U_k),
\end{equation}
where $\tilde\rho$ denotes a penalty function introduced to regularize
$\U_k$, and $\P^{+,k}_R \subseteq \P^{+}_R$ allows for constraints on
$\U_k$ (which may be different for each component of the mixture).
The exact penalties and constraints considered are described
below. When $\P^{+,k}_R = \P^{+}_R$ and $\tilde{\rho}(\U_k) = 0$ for
all $\U_k$, solving \eqref{eq:penloglik} corresponds to
maximum-likelihood estimation of $\bpi, \cU$.

\item Compute the posterior distribution for each $\btheta_j$ given
  $\bpi, \U$ estimated in the first stage:
\begin{align}
\label{eq:post}
p_{\mathsf{post}}({\bm\theta}_j) \colonequals&\;
p(\bs{\theta}_j \mid \x_j, \hat\bpi, \hat{\cU}) \nonumber \\
\propto&\; p(\x_j \mid \btheta_j) \, p(\btheta_j \mid \hat{\bpi}, \hat{\cU}).
\end{align}

\end{enumerate}
The posterior distributions \eqref{eq:post} have an analytic form, and
are mixtures of multivariate normal distributions (see for example
\citealt{urbut2019flexible, bovy2011extreme}) and Section
\ref{appendix:posterior_theta} of the Supplementary Materials;
\citealt{supplementary_text}). Since these posterior distributions
have a closed form, it is relatively straightforward to compute
posterior summaries such as the posterior mean and posterior standard
deviation, and measures for significance testing such as the {\em
  local false sign rate} ({\em lfsr}) \citep{stephens2017false}.
Therefore, we focus on methods for accomplishing the first step,
solving \eqref{eq:penloglik}.

\subsection{A reformulation to ensure scale invariance}

Intuitively, one might hope that changing the units of measurement of
all the observed $\x_j$ would simply result in corresponding changes
to the units of the estimated $\btheta_j$.  This idea can be
formalized as requiring that solutions of the EBMNM model should obey
a ``scale invariance'' property. This consideration motivates us to
reformulate \eqref{eq:penloglik}. Let $\hat{\bm{\theta}}_j(\x_1,
\ldots, \x_n, \V_1, \ldots, \V_n)$ denote the posterior mean for
$\bm{\theta}_j$ computed by solving the EBMNM problem with data $\x_1,
\ldots, \x_n, \V_1, \ldots, \V_n$. We say that the solution is ``scale
invariant'' if, for any $s > 0$, the following holds:
\begin{equation}
\label{eq:scale_invariant}
\hat{\bm{\theta}}_j(s\x_1, \ldots, s\x_n, s^2\V_1, \ldots, s^2\V_n)
= s\hat{\bm{\theta}}_j(\x_1, \ldots, \x_n, \V_1, \ldots, \V_n).
\end{equation}
That is, multiplying all the observed data points by $s$ (which
multiplies the corresponding error variance by $s^2$) has the effect
of multiplying the estimated means by $s$.  (Note: we state scale
invariance in terms of the posterior means only for simplicity; the
concept is easily generalized to require that the whole posterior
distribution for $\btheta_j$ scales similarly.)

Without the penalty $\tilde{\rho}$ in \eqref{eq:penloglik}, it is easy
to show that the scale invariance property \eqref{eq:scale_invariant}
holds provided that the constraints satisfy $\U_k \in \P^{+,k}_R
\implies s_k\U_k \in \P^{+,k}_R, \forall s_k > 0$. With penalty, scale
invariance holds provided that $\tilde{\rho}(\U) = \tilde{\rho}(s\U),
\forall s, \U$; that is, provided that the penalty function depends
only on the ``shape'' of $\U$ and not on its ``scale''.  To ensure
scale invariance, we therefore consider penalties of the form
\begin{equation}
\tilde{\rho}(\U) = \min_{s \,>\, 0} \, \rho(\U/s),
\label{eq:penalty-scale-invariant}
\end{equation}
where $\rho$ is a penalty that may depend on both the shape and scale
of $\U$. To give some intuition, suppose that $\rho$ encourages all
the eigenvalues of $\U$ to be close to 1. Then $\tilde{\rho}$ will
encourage the eigenvalues to be close to each other, without requiring
that they specifically be close to 1. Therefore, plugging
\eqref{eq:penalty-scale-invariant} into \eqref{eq:penloglik}, we have
\begin{equation}
(\hat{\bpi}, \hat{\cU})
\colonequals \argmax_{\bpi \,\in\, \cS_K, \, \U_k \,\in\, \P^{+,k}_R,\, 
\s \,>\, \bm{0}}
\ell(\bpi, \cU) - \sum_{k=1}^K \rho(\U_k/s_k),
\label{eq:penloglik2}
\end{equation}
where $\bm{s} \colonequals (s_1, \ldots, s_K)$. We use
\eqref{eq:penloglik2} for the remainder of the paper.

\subsection{Constraints and penalties}

Estimating covariance matrices in high-dimensional settings ({\em
  i.e.}, large $R$) is known to be a challenging problem (e.g.,
\citealt{johnstone2018, fan2016, ledoit2022}). Even in the simpler
setting of independent and identically distributed observations from a
single multivariate normal distribution, the maximum-likelihood
estimate of the covariance matrix can be unstable, and so various
covariance regularization approaches have been proposed to 
address this issue \citep{ledoit2004well, friedman2008sparse,
  cai2011adaptive, won2013condition, chi2014stable}.  Interestingly,
in the context of using EBMNM for significance testing, adding
penalties have additional benefits (see the numerical experiments below).

We consider two different penalties that have been previously used for
covariance regularization:
\begin{enumerate}
\setlength{\itemsep}{6pt}  
\item The ``inverse Wishart'' (IW) penalty 
\begin{align}
\label{eq:iwpen}
\rho_{\lambda}^{\mathrm{IW}}(\U) &\colonequals
\frac{\lambda}{2}\{\log|\U|
+ \mathrm{tr}({\U}^{-1})\} \\
& = \frac{\lambda}{2} \sum_{r=1}^R \big(\log e_r + 1/e_r\big).
\label{eq:iwpen_e}
\end{align}
\item The ``nuclear norm'' (NN) penalty:
\begin{align}
\label{eq:nupen}
\rho_{\lambda}^{\mathrm{NN}}(\U) &\colonequals
\frac{\lambda}{2}\{0.5\|\U\|_{\ast}+0.5\|\U^{-1}\|_{\ast}\} \\
&= \frac{\lambda}{2} \sum_{r=1}^R \big(0.5 e_r + 0.5/e_r\big).
\label{eq:nupen_e}
\end{align}
\end{enumerate}
Here, $e_1, \dots, e_R$ denote the eigenvalues of $\U$, and $\lambda >
0$ controls the strength of the penalty. We chose $\lambda = R$ in our
simulations, but one could also use cross-validation to select
$\lambda$. 
%
%

The IW penalty on $\U_k$ corresponds to {\it maximum a posteriori}
(MAP) estimation of $\U_k$ under an inverse-Wishart prior, with prior
mode $\I_R$ and $\lambda-R-1$ degrees of freedom
\citep{fraley2007bayesian}.  This penalty was also mentioned (but not
used or evaluated) in \cite{bovy2011extreme}.

The nuclear norm penalty was studied as an alternative to the IW
penalty for estimation of covariance matrices in
\cite{chi2014stable}. To our knowledge, this penalty has not been
studied in the EBMNM setting. The nuclear norm penalty in
\cite{chi2014stable} includes a hyperparameter $\alpha \in (0,1)$ that
controls the balance between $\|\U\|_{\ast}$ and $\|\U^{-1}\|_{\ast}$,
but our approach to ensuring scale-invariance has the consequence that
changing $\alpha$ is equivalent to changing $\lambda$ (see the
Supplementary Materials; \citealt{supplementary_text}), so we set
$\alpha = 0.5$.

As can be seen from \eqref{eq:iwpen_e} and \eqref{eq:nupen_e}, both
penalties depend only on the eigenvalues of $\U$, and decompose into
additive functions of the $R$ eigenvalues. Both penalties are
minimized when $\U = \I_R$, and more generally encourage $\U$ to be
well-conditioned by making it closer to the identity matrix (by
pushing the eigenvalues closer to 1).

As an alternative to penalized estimation of $\U$, we also consider
estimating $\U$ under different constraints:
\begin{enumerate}
\setlength{\itemsep}{6pt}  

\item A scaling constraint, $\U_k = c_k \U_{0k}$, for some chosen
  $\U_{0k} \in \P^{+}_R$.

\item A rank-1 constraint, $\U_k = \u_k \u_k^T$, for some $\u_k \in
  \mathbb{R}^R$.


\end{enumerate}
The scaling constraint could be useful for situations in which the
$\btheta_j$ may obey an expected sharing structure, or for sharing
structures that are easier to interpret. For example, $\U_{0k} = \I_R$
captures the situation in which all the effects $\theta_{j1}, \dots,
\theta_{jR}$ are independent, and $\U_{0k} = \one \one^T$ captures the
situation in which all the effects $\theta_{j1}, \dots, \theta_{jR}$
are equal.  Such covariances are referred to as ``canonical'' covariance
matrices in \cite{urbut2019flexible}.

The rank-1 constraint also leads to potentially more interpretable
covariance matrices, and can be thought of as a form of regularization
because low-rank matrices have fewer parameters to be estimated. It
may also allow for faster computations. \cite{urbut2019flexible} in fact
make extensive use of the rank-1 constraint. However, our results will
show that this constraint can cause problems for significance testing
and so may be better avoided in practice.

\section{Fitting the EBMNM model}
\label{sec:fitting}

We now describe three algorithms we have implemented for fitting
variations of the EBMNM model described above: the ED algorithm from
\citealt{bovy2011extreme}; an algorithm based on methods commonly used
in factor analysis (FA); and another based on the truncated eigenvalue
decomposition (TED).  Each of these algorithms applies to a subset of
EBMNM models (Table~\ref{tab:algo_overview}). In some situations, only
one algorithm can be applied; for example, only ED can handle
heteroskedastic variances with no constraints on $\U$. However, in
other settings all three algorithms can be applied (e.g.,
homoskedastic errors, no constraints on $\U$). Below, we empirically
assess the relative merits of the different algorithms in simulations
that model these different settings. See also Supplementary Tables
\ref{tab:comp_complex1} and \ref{tab:comp_complex2}
\citep{supplementary_text} for a comparison of the algorithms'
computational properties in the different settings.

\begin{table}[t]
\centering
\caption{\rm Summary of the EBMNM algorithms and the situations in which
  they apply. In this table we consider 4 variations of EBMNM: no
  constraint on $\U$ or a rank-1 constraint; and homoskedastic (hom.)
  errors (all $\V_j$ are the same) or heteroskedastic (het.) errors
  (one or more $\V_j$ differ).  Abbreviations used in this table are:
  ED = Extreme Deconvolution), FA = factor analysis, TED = truncated
  eigenvalue decomposition. A checkmark ($\checkmark$) indicates that
  the algorithm (ED, FA, TED) can be applied to the particular
  variation. Note that fitting $\U$ with a scaling constraint involves
  only a 1-d optimization and is treated separately.}
\begin{tabular}{rc@{\;\;}cc@{\;\;}c}
& \multicolumn{4}{c}{constraints on $\U$} \\
& \multicolumn{2}{c}{none} &
\multicolumn{2}{c}{rank-1} \\
\cmidrule(lr){2-3} \cmidrule(lr){4-5}
algorithm & hom. & het. & hom. & het. \\[0.75ex]
\rowcolor{Gray}
ED & $\checkmark$ & $\checkmark$ & & \\
FA & $\checkmark$ & & $\checkmark$ &
$\checkmark$ \\ 
\rowcolor{Gray}
TED & $\checkmark$ & & $\checkmark$ & 
\end{tabular}
\label{tab:algo_overview}
\end{table}

\subsection{Algorithms for the single-component EBMNM model with no penalty}

While these algorithms are implemented for the mixture prior
\eqref{eq:prior} with the penalties described above, the algorithms
are much easier to describe in the special case of one mixture
component ($K = 1$), and without penalty. So initially we focus on
this simpler case, and later we extend to the general form with
penalties and $K \geq 1$.

With $K = 1$ and no penalty, the prior is $\btheta_j \sim N(\zero,
\U)$, the model is
\begin{equation}
\label{eq:normal_prior_model}
\x_j \mid \U \sim N_R(\zero, \U+\V_j),
\quad j = 1, \dots, n.
\end{equation}
and the goal is to compute the maximum-likelihood estimate of $\U$:
\begin{equation}
\hat\U 
\colonequals \argmax_{\U \,\in\, \P_R^{+}} \sum_{j=1}^n 
\log N_R(\x_j; \zero, \U + \V_j)
\label{eq:simplemax}
\end{equation}

The three algorithms for solving \eqref{eq:simplemax} are as follows.

\paragraph*{Truncated Eigenvalue Decomposition (TED)}

This algorithm, which to the best of our knowledge is new in this
context, exploits the fact that, in the special case where $\V_j =
\I_R$, $j = 1, \ldots, n$, the maximum-likelihood estimate
\eqref{eq:simplemax} can be computed exactly. At first glance, one
might try to solve for $\hat\U$ by setting $\hat{\U} + \I_R$ to the
sample covariance matrix, $\S \colonequals \sum_{j=1}^n \x_j
\x_j^T/n$, then recovering $\hat{\U}$ as $\hat{\U} = \S -
\I_R$. However, $\S-\I_R$ is not necessarily a positive semi-definite
matrix; that is, it may have one or more eigenvalues that are
negative. One could deal this problem by setting the negative
eigenvalues to zero, and indeed this approach is correct; that is,
letting $(\S)_{+}$ be the matrix obtained from $\S$ by truncating its
negative eigenvalues to 0, $\hat\U=(\S-\I_R)_+$ is the
maximum-likelihood estimate of $\U$ \citep{tipping1999probabilistic}.
This idea can also be used to solve the more general case, $\V_j =
\V$, essentially by transforming the data to $\R^{-1}\x_j$ where $\V=\R\R^T$, estimating $\R^{-1} \U \R^{-T}$ from this transformed data, then
reversing this transformation, see \citealt{supplementary_text}. 

\paragraph*{Extreme Deconvolution (ED)}

This is an EM algorithm \citep{dempster1977maximum}, an iterative
approach to solving \eqref{eq:simplemax}; the name comes from
\cite{bovy2011extreme}.  ED uses the natural ``data augmentation''
representation of \eqref{eq:normal_prior_model}:
\begin{equation}
\label{eq:ed_aug}
\begin{aligned}
\btheta_j &\sim N_R(\zero, \U) \\
\x_j \mid \btheta_j &\sim N_R(\btheta_j, \V_j).
\end{aligned}
\end{equation} 
Following the usual EM derivation, the updates can be derived as 
\begin{equation}
\label{eq:simple_ed_update_U_explicit}
\U^{\mathrm{new}} = \frac{1}{n} \sum_{j=1}^n \bs{B}_j+\bs{b}_j\bs{b}_j^T,
\end{equation}
where $\bm{b}_j$ and $\bm{B}_j$ are, respectively, the posterior mean
and covariance of $\btheta_j$ given $\U$:
\begin{align}
\bs{b}_j &\colonequals \U(\U+\V_j)^{-1}\bs{x}_j \label{eq:conditional_b} \\
\bs{B}_j &\colonequals \U-\U(\U+\V_j)^{-1}\U. \label{eq:conditional_B}
\end{align}
The update \eqref{eq:simple_ed_update_U_explicit} is guaranteed to
increase (or not decrease) the objective function in
\eqref{eq:simplemax}, and repeated application of
(\ref{eq:simple_ed_update_U_explicit}--\ref{eq:conditional_B}) will
converge to a stationary point of the objective.

\paragraph*{Factor analysis (FA)}

This is also an EM algorithm, but based on a different data
augmentation than ED; the name comes from its close connection to EM
algorithms for factor analysis models \citep{ghahramani1996algorithm,
  rubin1982algorithms, zhao2008ml, liu1998maximum,
  mixture_models_book}. In its simplest form, the FA approach imposes
a rank-1 constraint on $\U$, $\U = \u\u^T$, where $\u \in
\mathbb{R}^R$ is to be estimated.  The model
\eqref{eq:normal_prior_model} then admits the following data
augmentation representation:
\begin{equation}
\label{eq:fa_aug}
\begin{aligned}
a_j &\sim N(0,1) \\
\x_j \mid \u, \V_j, a_j &\sim N(a_j \u, \V_j).
\end{aligned}
\end{equation}
The usual EM derivation gives the update 
\begin{equation}
\label{eq:simple_u_fa}
\u^{\mathrm{new}} =
\bigg(\sum_{j = 1}^n (\mu_{j}^2 + \sigma_{j}^2) \V_j^{-1}\bigg)^{-1}
\bigg(\sum_{j = 1}^n \mu_j \V_j^{-1} \x_j\bigg),
\end{equation}
in which $\mu_j$ and $\sigma_j^2$ denote, respectively, the posterior
mean and posterior covariance of $a_j$ given $\u$,
\begin{align}
\label{eq:mu}
\mu_{j} &\colonequals \sigma_{j}^2\u^T\V_j^{-1}\x_j \\
\label{eq:sigma}
\sigma_{j}^2 &\colonequals 1/(1 + \u^T\V_j^{-1}\u). 
\end{align}
The update \eqref{eq:simple_u_fa} is guaranteed to increase (or not
decrease) the objective function in \eqref{eq:simplemax}, and repeated
application of (\ref{eq:simple_u_fa}--\ref{eq:sigma}) will converge to
a stationary point of the objective.  These updates can be extended to
higher-rank covariances where the goal is to find the
maximum-likelihood estimate subject of $\U$ subject to $\U$ having
rank at most $R'$, where $R' \leq R$. However, when $R'>1$, the
updates are have closed-form expressions only for homoskedastic
errors, $\V_j = \V$.

The three algorithms have different strengths and weaknesses, and
settings to which they apply (Table \ref{tab:algo_overview}).  The TED
algorithm has the advantage of directly computing the
maximum-likelihood estimate, which seems preferable to an iterative
approach. However, TED only applies in the case of homoskedastic
errors. The ED approach is more general, applying to both
heteroskedastic and homoskedastic errors, although it cannot fit
rank-1 matrices. The FA approach is attractive for low-rank matrices,
particularly for fitting rank-1 matrices.

Our claim that ED cannot fit rank-1 covariance matrices deserves
discussion, especially since \cite{urbut2019flexible} used ED to fit
such matrices.  As pointed out by \cite{urbut2019flexible}, if ED is
initialized to a low-rank matrix with rank $R'$, then the updated
estimates \eqref{eq:simple_ed_update_U_explicit} are also rank (at
most) $R'$. Thus, if ED is initalized to a rank-1 matrix, the final
estimate is also rank-1. However, the ED estimates are not only low
rank, but also span the same subspace as the initial estimates, a
property we refer to as ``subspace-preserving''. Thus, if ED is
initialized to $\U = \u \u^T$, the updated estimates will be of the
form $a \u \u^T$ for some scalar $a$. In other words, the ED update
does not change rank-1 matrices, except by a scaling factor, and so
the final estimate will simply be proportional to the initial
estimate. This flaw motivated us to implement the FA method. However,
as our numerical comparisons will illustrate, the rank-1 matrices turn
out to have other drawbacks that lead us not to recommend their use
anyhow.

\subsection{Extending the algorithms to a mixture prior}

\begin{algorithm}[t]
\SetAlgoNoLine
\caption{EM for fitting the EBMNM model.}
\label{algo:general_EM}

\KwIn{Data vectors $\x_j \in \mathbb{R}^{R}$ and correspoding
  covariance matrices $\V_j \in \P_R^{+}$, $j = 1, \ldots, n$; $K$,
  the number of mixture components; initial estimates of the prior
  covariance matrices $\cU = \{\U_1, \ldots, \U_K\}$, $\U_k \in
  \P_R^{+,k}$, $k = 1, \ldots, K$; initial estimates of the scaling
  parameters $\bm{s} = \{s_1, \ldots, s_K\} \in \mathbb{R}^K$; 
  initial estimates of the mixture weights $\bpi \in \cS_K$.}

\KwOut{$\cU$, $\bpi$.}

\Repeat{\rm convergence criterion is met} {

  \For{$j \leftarrow 1$ \KwTo $n$}{
    \For{$k \leftarrow 1$ \KwTo $K$}{
      Update $w_{jk}$ using \eqref{eq:responsibilities}.
    }
  }

   \For{$k \leftarrow 1$ \KwTo $K$} {
     
     $\pi_k \leftarrow \sum_{j=1}^n w_{jk}/n$
     
     $\U_k\leftarrow\argmax_{\U \,\in\, P_R^{+,k}}
     \phi(\U; \w_k) - \rho(\U/s_k)$

     \hspace{2em} $\triangleright$ Note that some algorithms compute
     this argmax inexactly.
     
     $s_k \leftarrow \argmin_{s\,>\,0} \rho(\U_k/s)$
   }
}
\end{algorithm}

All of the algorithms---TED, ED and FA---can be generalized from the
$K = 1$ case to the $K \geq 1$ case using the standard EM approach to
dealing with mixtures.  The resulting algorithms have a simple common
structure, summarized in Algorithm \ref{algo:general_EM}. (This
algorithm also allows for an inclusion of a penalty, which is treated
in the next section. See also the Supplementary Materials for a
derivation of this algorithm; \citealt{supplementary_text}.) This EM
algorithm involves iterating the following steps: (i) compute the
weights, $w_{jk}$ (sometimes called the ``responsibilities''), each
which represent the conditional probability that mixture component $k$
gave rise to observation $j$ given the current estimates of $\bpi,
\cU$,
\begin{equation}
w_{jk} =
\frac{\pi_k N_R(\x_j; \zero, \U_k + \V_j)}{
\sum_{k'=1}^K \pi_{k'} N_R(\x_j; \zero, \U_{k'} + \V_j)};
\label{eq:responsibilities}
\end{equation}
(ii) update $\bpi$ by averaging the weights (this is the standard EM
update for estimating mixture proportions, and is the same for all the
algorithms); (iii) update the covariance matrices $\cU$ (this is the
step where the algorithms differ); and (iv) update the scaling
parameters, $\s$. (The update of the scaling parameters is the same
for all algorithms, and depends on the chosen penalty. For details,
see the Supplementary Materials; \citealt{supplementary_text}.)


With this data augmentation, the updates for the covariance matrices
$\U_k$ have the following form:
\begin{equation}
\U^{\mathrm{new}} = \argmax_{\U \,\in\, \P^{+,k}_R}
\phi(\U; \w_k),
\label{eq:maxh_no_penalty}
\end{equation}
where
\begin{equation} 
\phi(\U; \w) \colonequals \sum_{j=1}^n w_j
\log N_R(\x_j; \zero, \U + \V_j).
\label{eq:h}
\end{equation}
The function $\phi$ can be viewed as a {\it weighted} version of the
log-likelihood with normal prior \eqref{eq:normal_prior_model}, and
the updates for this weighted problem are very similar to the updates
for a normal prior. For example, the TED update, which solves the
weighted problem exactly in the case $\V_j = \I_R$, involves
truncating the eigenvalues of $\hat\S-\I_R$ where $\hat{\bs{S}}
\colonequals \sum_{j=1}^n w_j \x_j \x_j^T/(\sum_{j=1}^n w_j)$ is {\em
  the weighted sample covariance matrix.} Details of the TED, ED and
FA updates for weighted log-likelihoods are given in the Supplementary
Text \citep{supplementary_text}.

\subsection{Modifications to the algorithms to incorporate the penalties}

With a penalty, the updates \eqref{eq:maxh_no_penalty} are instead
\begin{equation}
\U^{\mathrm{new}} = \argmax_{\U \,\in\, \P^{+,k}_R, \, s_k \,>\, 0}
\phi(\U; \w_k) - \rho(\U/s_k).
\label{eq:maxh}
\end{equation}

We have adapted the TED approach, in the case $\V_j =\I_R$, to
incorporate either the IW or NN penalty. These extensions replace the
simple truncation of the eigenvalues with solving a (1-d) optimization
problem for each eigenvalue $e_r$; while these 1-d optimization
problems do not have closed form solutions, they are easily solved
using off-the-shelf numerical methods. This approach can also be
applied, via the data transformation approach, to the general
homoskedastic case $\V_j =\V$; however, this transformation approach
implicitly changes the penalty so that it encourages $\U_k/s_k$ to be
close to $\V$ rather than being close to $\I_R$. This change in the
penalty appears to be necessary to make the TED approach work when $\V
\neq \I_R$. See the Supplementary Materials \citep{supplementary_text}
for details.

Incorporating an IW penalty into the ED approach is also
straightforward \citep{bovy2011extreme} and results in a simple
change to the closed-form updates
\eqref{eq:simple_ed_update_U_explicit}. The NN penalty does not result
in closed-form updates for ED and so we have not implemented it.

Incorporating penalties into the FA updates may be possible but we
have not done so.

\subsection{Significance testing}
\label{sec:lfsr}

In EBMNM, inferences are based on the posterior distribution,
$p_{\mathsf{post}}(\btheta_j) = p(\btheta_j \mid \x_j, \hat{\cU},
\hat{\bpi})$, in which $\hat{\cU}, \hat{\bpi}$ denote the estimates
returned by Algorithm~\ref{algo:general_EM}.  To test for
significance, we use the local false sign rate ({\em lfsr}), which has
been used in both univariate \citep{stephens2017false,
  xie2002discussion} and multivariate \citep{urbut2019flexible,
  liu2022flexible, mvsusie} settings. The {\em lfsr} is defined as
\begin{equation}
\mbox{\em lfsr}_{jr} \colonequals
\min\{p_{\mathsf{post}}(\theta_{jr}\ge 0),
      p_{\mathsf{post}}(\theta_{jr}\le 0)\}.
\label{eq:lfsr}
\end{equation}
In particular, a small {\em lfsr} indicates high confidence in the
sign of $\theta_{jr}$. The {\em lfsr} is robust to modeling
assumptions, which is helpful for reducing sensitivity to the choice
of prior \citep{stephens2017false}.

\section{Numerical comparisons}
\label{sec:simulation}


We ran simulations to (i) compare the performance of the different
approaches to updating the covariance matrices ${\bm U}_k$; (ii)
assess the benefits of the penalties and constraints; and (iii) assess
the sensitivity of the results to the choice of $K$, the number of
mixture components in the prior. 

We used the Dynamic Statistical Comparisons software
(\url{https://github.com/stephenslab/dsc}) to perform the
simulations. Code implementing the simulations is provided in the
supplementary ZIP file \citep{supplementary_zip}, and includes a
workflowr website \citep{workflowr} for browsing the results. The code
and workflowr website is also available online at
\url{https://github.com/yunqiyang0215/udr-paper}.


\subsection{Data generation}

We simulated all data sets from the EBMNM model
(\ref{eq:mvnm}--\ref{eq:prior}); that is, for each data set, we
simulated the ``true'' means $\btheta_1, \ldots, \btheta_n \in \mathbb{R}^R$
from the mixture prior \eqref{eq:prior} then we simulated observed
data, the ``noisy'' means $\x_1, \ldots, \x_n \in \mathbb{R}^R$,
independently given the true means. Note that model fitting and
inferences were performed using only $\x_1, \ldots, \x_n$; the true
means $\btheta_1, \ldots, \btheta_n$ were only used to evaluate
accuracy of the inferences.  Further, to evaluate the ability of the model
to generalize to other data, we also simulated test sets with true
means $\btheta_1^{\mathrm{test}}, \ldots,
\btheta_{n_{\mathrm{test}}}^{\mathrm{test}}$ and observed vectors
$\x_1^{\mathrm{test}}, \ldots,
\x_{n_{\mathrm{test}}}^{\mathrm{test}}$. These test sets were
simulated in the same way as the training sets.

In all cases, we set the number of mixture components $K$ to 10, with
uniform mixture weights, $\pi_1, \ldots, \pi_{10} = 1/10$. We
generated the $K = 10$ covariance matrices $\U_1, \ldots, \U_{10}$ in two
different ways, which we refer to as the ``hybrid'' and ``rank-1''
scenarios:
\begin{enumerate}
\setlength{\itemsep}{6pt}  

\item {\bf Hybrid scenario.} We used 3 canonical covariance matrices,
  and randomly generated an additional 7 covariance matrices randomly
  from an inverse-Wishart distribution with scale matrix $\S = 5\I_R$
  and $\nu = R + 2$ degrees of freedom. The 3 canonical matrices were
  as follows: $\U_1 = 5\e_1 \e_1^T$, a matrix of all zeros except for
  a 5 in the top-left position, which generates ``singleton'' mean
  vectors with a single non-zero element, $\btheta_j = (\theta_{j1},
  0, \ldots, 0)$; $\U_2 = 5\one \one^T$, which generates equal means
  $\btheta_j = (\alpha_j, \ldots, \alpha_j)$ for some scalar
  $\alpha_j$; and $\U_3 = 5\I_R$, which generates mean vectors
  $\btheta_j$ that are independent in each dimension.
%
%
%
\item {\bf Rank-1 scenario.} We used 5 covariance matrices of the form
  $\U_k = 5 \e_k \e_k^T$, $k = 1, \ldots 5$, which generate mean
  vectors of length $R$ with zeros everywhere except at the $k$th
  position. The remaining 5 covariances were random rank-1 matrices of
  the form $\U_k = \u_k \u_k^T$, $\u_k \sim N_R(\bm{0},\I_R)$,
  $k=6,\dots,10$.
  
\end{enumerate}

In both scenarios, we simulated large, low-dimension data sets ($n =
\mbox{10,000}$, $R = 5$), and smaller, high-dimension data sets ($n =
\mbox{1,000}$, $R = 50$). We refer to these data sets as ``large $n/R$''
and ``small $n/R$'', respectively.  To allow comparisons between TED, ED
and FA, in all cases we simulated data sets with homoskedastic errors;
that is, $\V_j = \I_R$, $j = 1, \ldots, n$ and $\V_j^{\mathrm{test}} =
\I_R$, $j = 1, \ldots, n_{\mathrm{test}}$.

%
%

\subsection{Results}

\subsubsection{Comparison of convergence}

We first focused on comparing the convergence of the different updates
(TED, ED and FA). For brevity, we write ``TED'' as shorthand for ``the
algorithm with TED updates'', and similarly for ED and FA. To compare
the updates under the same conditions, we used FA here to fit
full-rank covariance matrices, not rank-1 matrices. We ran TED and ED
with and without the IW penalty.


\begin{figure}[t]
\centering
\includegraphics[width=0.95\textwidth]{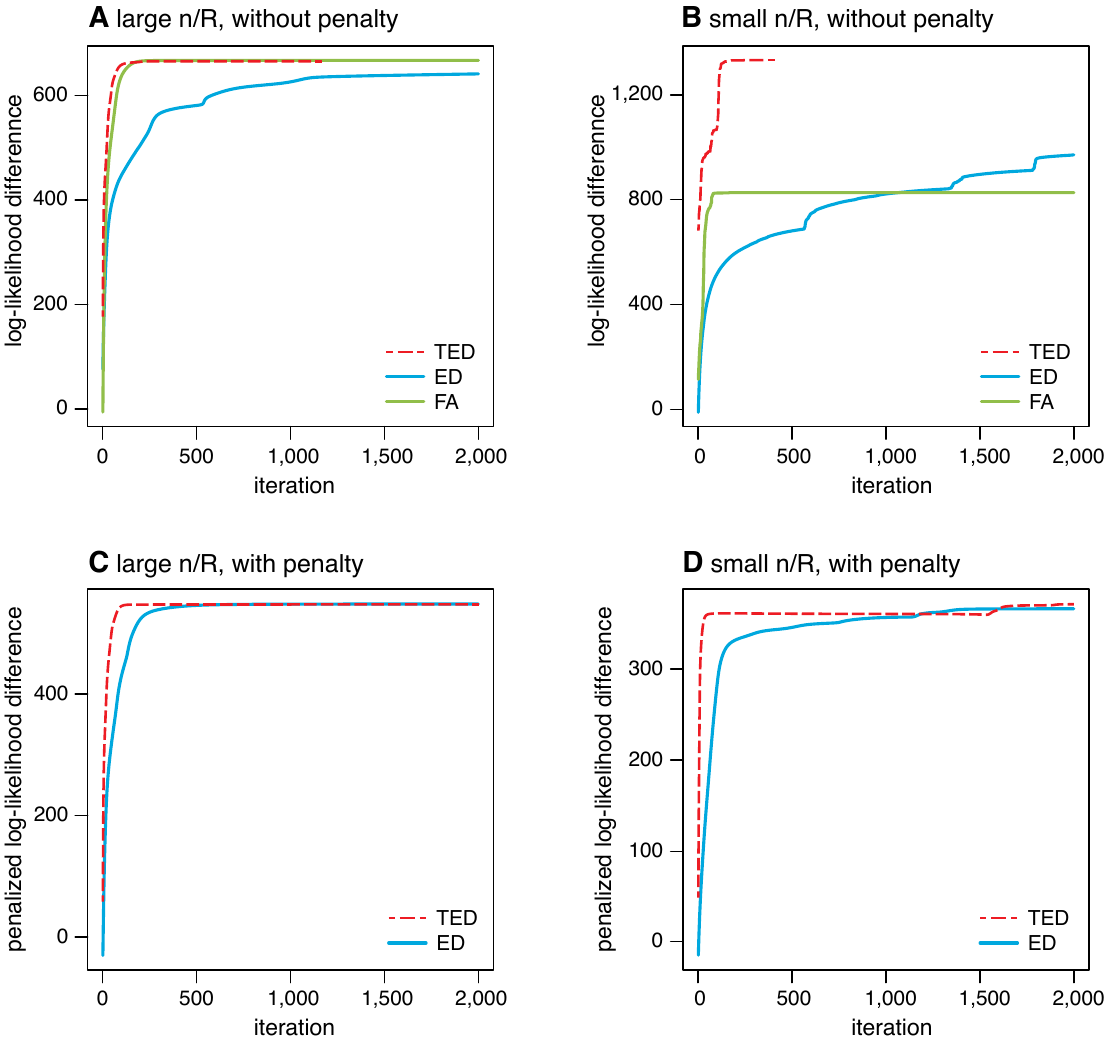}
\caption{\rm Illustrative examples comparing convergence of TED, ED
  and FA. Each plot shows the algorithms' progress over iterations on
  a single simulated data set. A and C show the results for the same
  data set, and similarly for B and D. In all cases, we ran 2,000
  iterations, although in some cases, TED updates stopped early
  because the updates converged to a stationary point of the objective
  (the likelihood or the penalized likelihood). For examples A and B,
  no penalty was used; for examples C and D, the inverse Wishart (IW)
  penalty was used with penalty strength $\lambda = R$. Log-likelihood
  and penalized log-likelihood differences are plotted with respect to
  the (penalized) log-likelihood near the initial estimate. An initial
  round of 20 ED iterations common to all the runs (the ``warm
  start'') is not shown in these plots.}
\label{fig:example}
\end{figure}

\begin{figure}[t]
\centering
\includegraphics[width=0.825\textwidth]{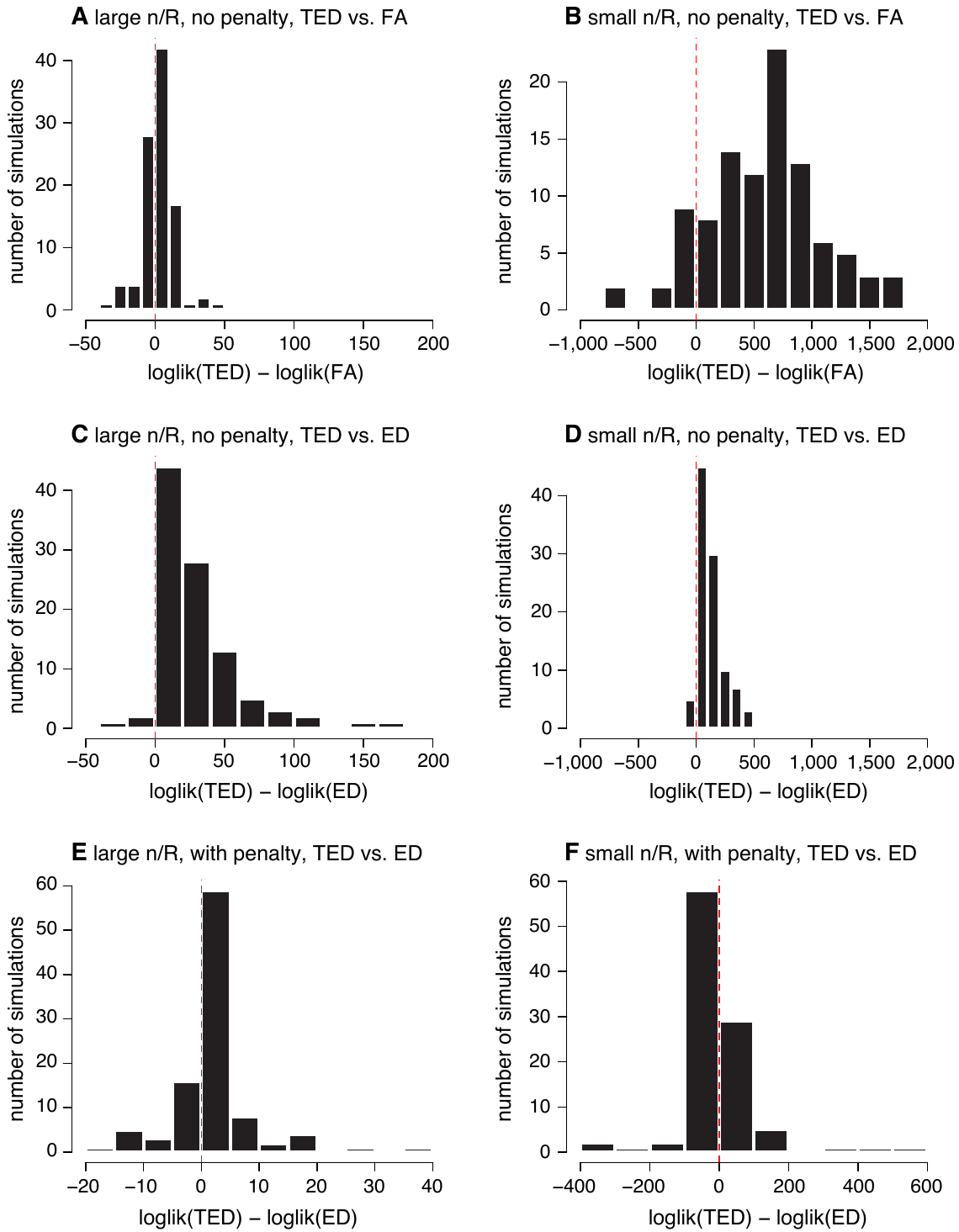}
\caption{\rm Comparison of convergence of TED, ED and FA. Each plot
  summarizes the results from 100 simulations. In each simulation,
  log-likelihood achieved after (at most) 2,000 iterations was
  recorded. Panels E and F show differences in the penalized
  log-likelihoods (IW penalty, $\lambda = R$).}
\label{fig:histdiff}
\end{figure}

We ran all methods on 100 ``large $n/R$'' data sets and 100 ``small
$n/R$'' data sets simulated in the hybrid scenario, setting $K=10$ to
match the simulated truth. To reduce the likelihood that the different
updates converge to different local solutions, we performed a
prefitting stage in which we ran 20 iterations of ED from a random
initial starting point (specifically, the initialization was $\pi_k =
1/10$, $s_k = 1$, with randomly generated $\U_k$, $k = 1, \ldots,
10$). We call this a ``warm start''. We then ran each algorithm for at
most 2,000 iterations after this warm start. Figure \ref{fig:example}
shows illustrative results for two data sets (one large $n/R$ and one
small $n/R$), and Figure \ref{fig:histdiff} summarizes the results
across all simulations.

The results in Figure \ref{fig:example} illustrate typical
behavior. Among the unpenalized updates, TED and FA converged much
more quickly than ED. For the penalized updates, TED still converged
more quickly than ED, but the difference is less striking than without
the penalty. In the small $n/R$ example the three methods appear to have
converged to different solutions (despite the use of a warm start).
This is not unexpected due to the non-convexity of the objective
function, and illustrates an important general point to keep in mind:
improvements in the quality of solution obtained may be due to either
faster convergence to the solution, convergence to a local solution
with higher objective, or both.

The results in Figure \ref{fig:histdiff} confirm that some of the
patterns observed in the illustrative example are true more generally.
In the unpenalized case (Panels A--D), TED often arrived at a better
solution than FA or ED, although in the large $n/R$ setting FA was
comparable to TED (Panel A). In the penalized case, the TED and ED
solutions were much more similar for both large and small $n/R$ (Panels
E, F).



\begin{figure}[t]
\centering
\includegraphics[width=0.9\textwidth]{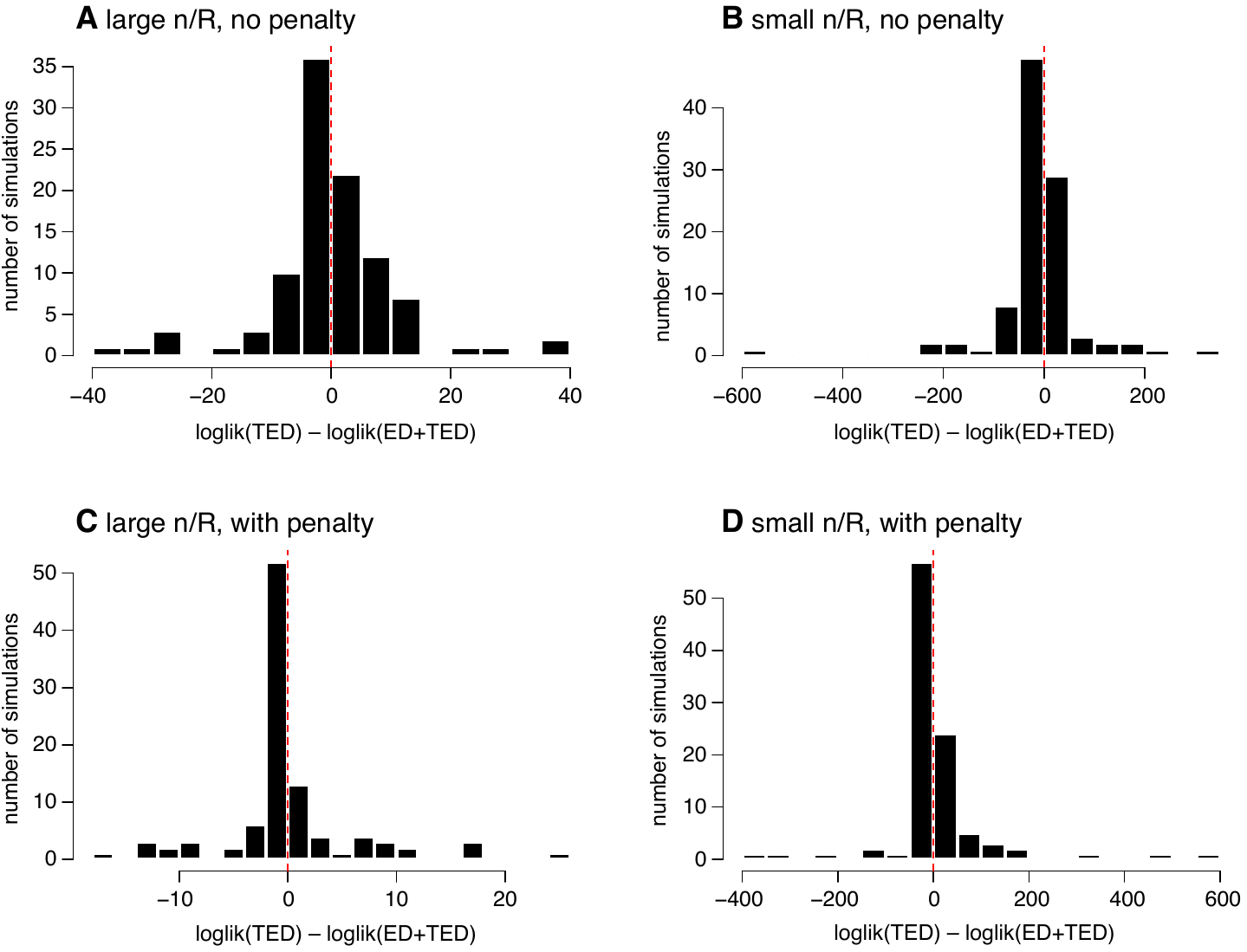}
\caption{\rm Results demonstrating the ability of TED to ``rescue''
  ED.  The log-likelihood or penalized log-likelihood obtained by
  performing at most 2,000 TED updates is compared against performing
  2,000 iterations of ED, followed by another round of (at most) 2,000
  iterations of TED updates (``ED+TED''). Each plot summarizes the
  results from 100 simulations, the same simulations as in
  Figure~\ref{fig:histdiff}.}
\label{fig:histdiff2}
\end{figure}

The improved performance of (unpenalized) TED vs. ED could be due
either to faster convergence (e.g., Figure \ref{fig:example}A) or due
to convergence to a better local optimum (e.g., Figure
\ref{fig:example}B). To investigate this, we assessed whether TED can
``rescue'' ED by running TED initialized to the ED solution
(``ED+TED'').
If ED converges to a poorer local optimum, then TED will not rescue
it, and ED+TED will be similar to ED; on the other hand, if ED is
simply slow to converge, then ED+TED will be similar to TED. The
results in Figure \ref{fig:histdiff2}
show that TED usually rescues ED; the ED+TED estimates were
consistently very similar to the TED estimates, regardless of whether
a penalty was used or not.  This suggests that the improved
performance of TED is generally due to faster convergence.

To assess how additional iterations improve ED's performance, we reran
ED for 100,000 iterations instead of only 2,000. Since these runs take
a long time, we examined only a few simulated data sets randomly
selected from the large $n/R$ and small $n/R$ scenarios (Supplementary
Figure~\ref{supple:fig1_convergence}). Even after 100,000 iterations,
ED fell measurably short of 1,000 updates of TED (average difference
of 40.6 log-likelihood units).

In summary, our experiments confirm that, for homoskedastic errors
(which is the setting where all three methods apply), TED exhibited
the best performance. This is not unexpected since TED solves the
subproblem (eq.~\ref{eq:maxh_no_penalty} or \ref{eq:maxh}) exactly
whereas ED and FA do not. Our experiments also show that including a
penalty can help improve convergence behavior, especially for
ED. Thus, including the penalty has computational benefits in addition
to statistical benefits demonstrated below.


\subsubsection{Comparison of the penalties and constraints}

Next we evaluated the benefits of different penalties and constraints
for estimation and significance testing of the underlying means
$\btheta_j$. We focused on TED and ED with and without penalties, and
on fitting rank-1 matrices using TED. (Since TED solves the subproblem
exactly, rather than iteratively, TED should be better than FA in this
setting with homoskedastic variances. The main benefit of FA is that
it can fit rank-1 matrices with heteroskedastic variances.)

We compared methods using the following evaluation measures:
\begin{itemize}
\setlength{\itemsep}{6pt}  

\item {\bf Plots of power vs. false sign rate (FSR).} These plots are
  similar to the more commonly-used plots of ``power vs. false
  discovery rate,'' but improve robustness and generality by requiring
  ``true discoveries'' to have the correct sign. The better methods are
  those that achieve higher power at a given FSR. See the
  Supplementary Text \citep{supplementary_text} for definitions.
  
\item {\bf Empirical False Sign Rate (FSR).} We report the empirical
  FSR among tests that were significant at $\text{\em lfsr} < 0.05$. A
  well-behaved method should have a small empirical FSR, certainly
  smaller than 0.05. We consider an FSR exceeding 0.05 to be
  indicative of a poorly behaved method.

\item {\bf Accuracy of predictive distribution.} To assess
  generalizability of the estimates of $\cU$ and $\bpi$, we compared
  the marginal predictive density \eqref{eq:marginal} in test samples,
  $p(\x_j^{\mathrm{test}} \mid \hat{\cU}, \hat{\bm\pi},
  \V_j^{\mathrm{test}})$, against the ``ground-truth'' marginal
  predictive density $p(\x_j^{\mathrm{test}} \mid \cU^{\mathrm{true}},
  {\bm\pi}^{\mathrm{true}}, \V_j^{\mathrm{test}})$, where
  $\cU^{\mathrm{true}}, {\bm\pi}^{\mathrm{true}}$ denote the
  parameters used to simulate the data. We summarized the relative
  accuracy in the predictions as
\begin{equation}
\frac{1}{n_\mathrm{test}}
\sum_{j=1}^{n_{\mathrm{test}}}
\log\bigg\{\frac{
p(\x_j^{\mathrm{test}} \mid \cU^{\mathrm{true}}, \bpi^{\mathrm{true}})}
{p(\x_j^{\mathrm{test}} \mid \hat{\cU}, \hat{\bpi})}\bigg\}.
\end{equation}
This measure can be interpreted as (an approximation of) the
Kullback-Leibler (K-L) divergence from the true predictive
distribution to the estimated predictive distribution. Smaller K-L
divergences are better.

\end{itemize}

\begin{figure}[t]
\centering
\includegraphics[width=0.925\textwidth]{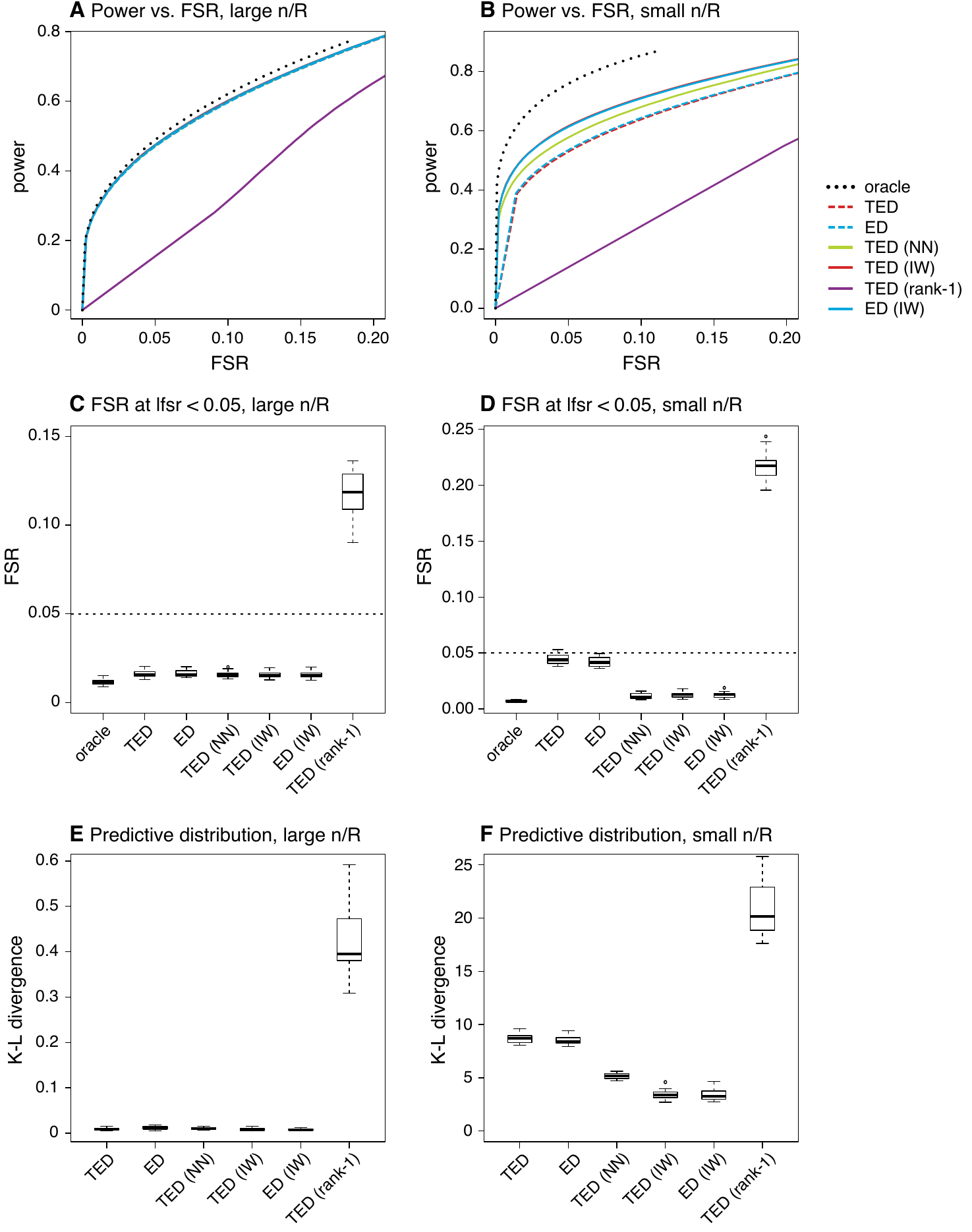}
\caption{\rm Comparison of penalties, constraints and updates (ED
  vs. TED) in the ``hybrid'' simulated data sets. For the IW and NN
  penalties, the penalty strength was set to $\lambda = R$. 
  In C and D, the target FSR is shown as a dotted horizontal line at
  0.05.  In A, most of the methods are not visible because the lines
  overlap at the top near the oracle result. Note that the oracle
  model always achieves a K-L divergence of zero.}
\label{fig:inference_general}
\end{figure}

\begin{figure}[t]
\centering
\includegraphics[width=0.95\textwidth]{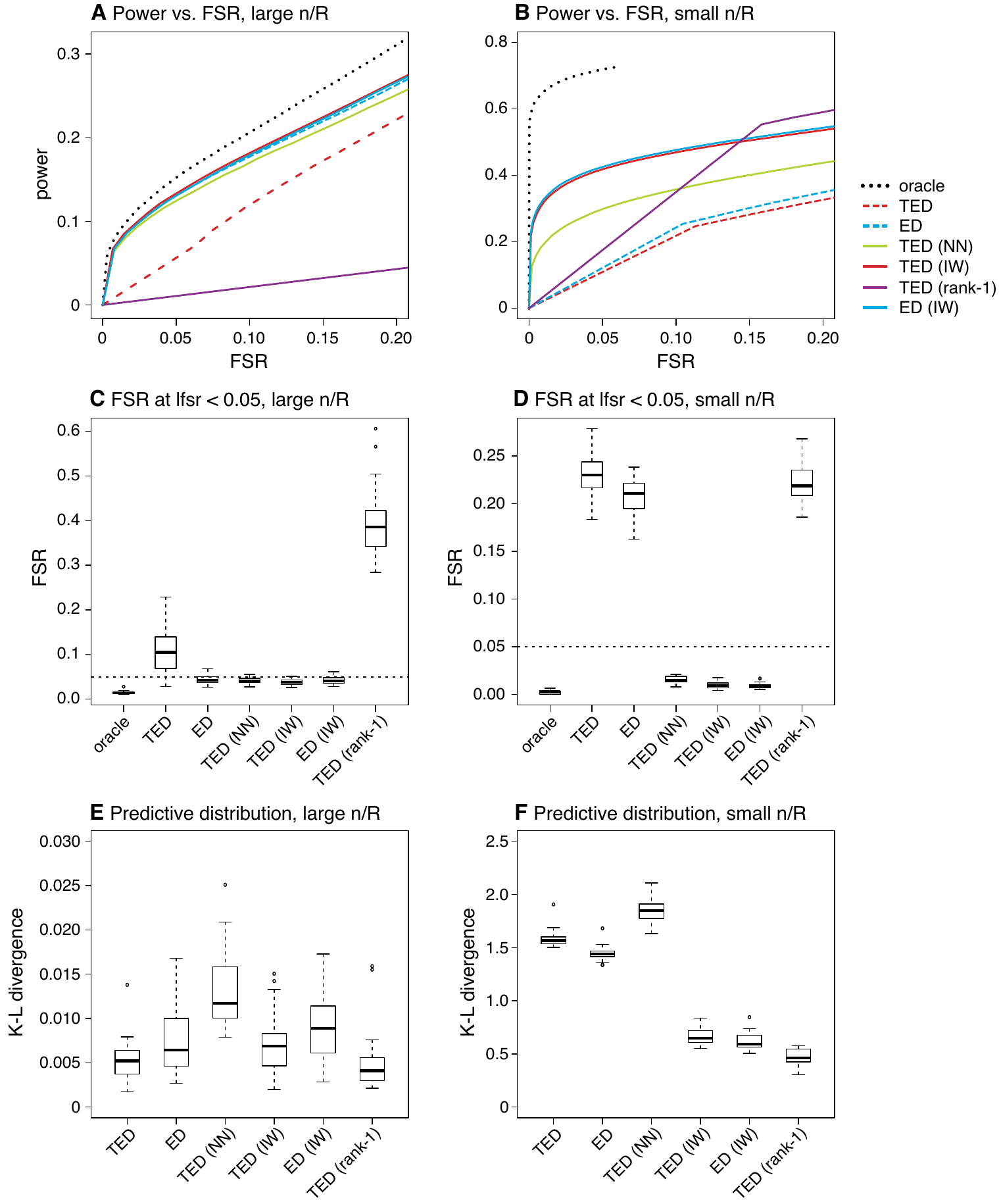}
\caption{\rm Comparison of penalties, constraints and updates (ED
  vs.~TED) in the ``rank-1'' simulated data sets. For the IW and NN
  penalties, the penalty strength was set to $\lambda = R$. 
  In C and D, the target FSR is shown as a dotted horizontal line at
  0.05. Note that the oracle model always achieves a K-L divergence of
  zero.}
\label{fig:inference_rank1}
\end{figure}


We simulated 20 data sets with large $n/R$ and another 20 data sets
with small $n/R$ under both the hybrid and rank-1 scenarios (80 data
sets in total). In all cases, model fitting was performed as above,
again with $K = 10$. The results are summarized in Figures
\ref{fig:inference_general} and \ref{fig:inference_rank1}.  In all
comparisons, we included results from the ``oracle'' EBMNM
model---that is, the model used to simulate the data---as a point of
reference.

Results for the hybrid setting are shown in Figure
\ref{fig:inference_general}. The results show a clear benefit of using
a penalty in the small $n/R$ setting: both IW and NN penalties improved
the power vs. FSR and the accuracy of predictive distributions.
For large $n/R$ data sets, the penalties do not provide a clear
benefit, but also do not hurt performance. In both cases TED and ED
perform similarly, suggesting that the poor convergence of ED observed
in previous comparisons may have less impact on performance than might
have been expected.  The rank-1 constraints performed very poorly in
all tasks, which is perhaps unsurprising since the true covariances
were (mostly) not rank-1.
%
%

%

Results for the rank-1 scenario are shown in
Figure~\ref{fig:inference_rank1}. In this case, imposing rank-1
constraints on the covariance matrices improved predictive
performance---which makes sense because the true covariances were
indeed rank-1---but produced worse performance in other metrics. In
particular, the {\em lfsr} values from the rank-1 constraint are very
poorly calibrated. This is because, as noted in
\cite{liu2022flexible}, the rank-1 constraint leads to {\em lfsr}
values that do not differ across conditions; see the Discussion
(Section~\ref{sec:discussion}) for more on this.  Penalized estimation
of the covariance matrices (using either a IW or NN penalty)
consistently achieved the best power at a given FSR in both the large
$n/R$ and small $n/R$ settings. Interestingly, (unpenalized) TED
performed much worse than (unpenalized) ED in the power vs. FSR for
large $n/R$. We speculate that this was due to the slow convergence of
ED providing sort of ``implicit regularization''. However, explicit
regularization via a penalty seems preferable to implicit
regularization via a poorly converging algorithm, and overall
penalized estimation was the winning (or equally winning) strategy
across a variety of settings.



 

\subsubsection{Robustness to mis-specifying the number of mixture components}

In the above experiments, we fit all models with a value of $K$ that
matched the model used to simulate the data. In practice, however, $K$
is unknown, and so we must also consider situations in which 
$K$ is mis-specified.
%
%
Intuitively, one might expect that overstating $K$ may lead to
overfitting and worse performance; \cite{urbut2019flexible} argued
however that the use of the mixture components centered at zero in the
prior \eqref{eq:prior} makes it robust to overfitting because the
mean-zero constraint limits their flexibility. Here we assess this
claim.
%
%

\begin{figure}[t]
\centering
\includegraphics[width=0.9\textwidth]{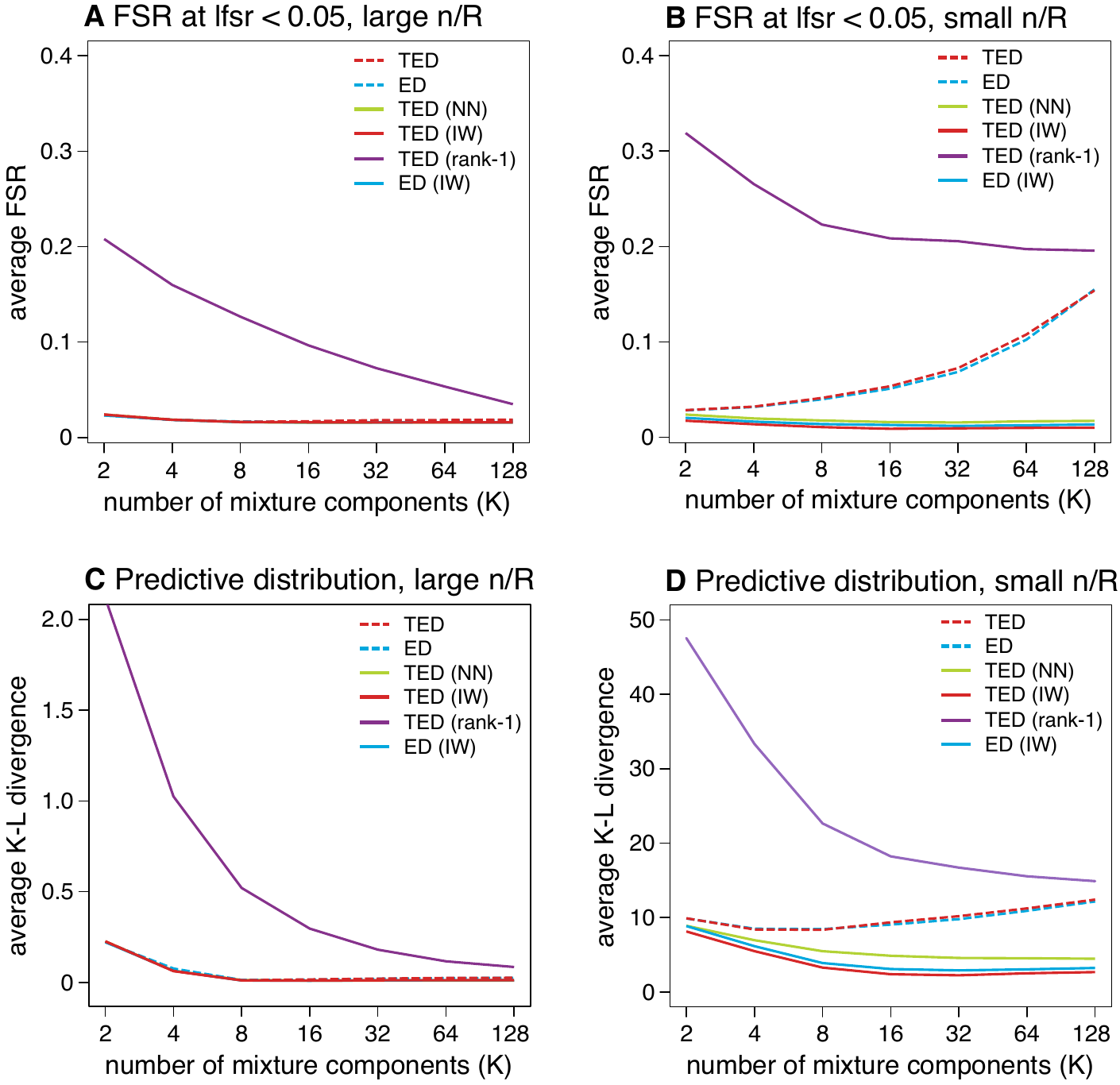}
\caption{\rm Assessment of robustness to mis-specifying $K$ in
  ``hybrid'' simulated data sets.
  All results are averages over 20 data sets, each simulated with $K =
  10$ mixture components. In A and C, most of the methods are not visible
  because they overlap with the ``TED-IW'' result.}
\label{fig:robustK}
\end{figure}

\begin{figure}[t]
\centering
\includegraphics[width=0.9\textwidth]{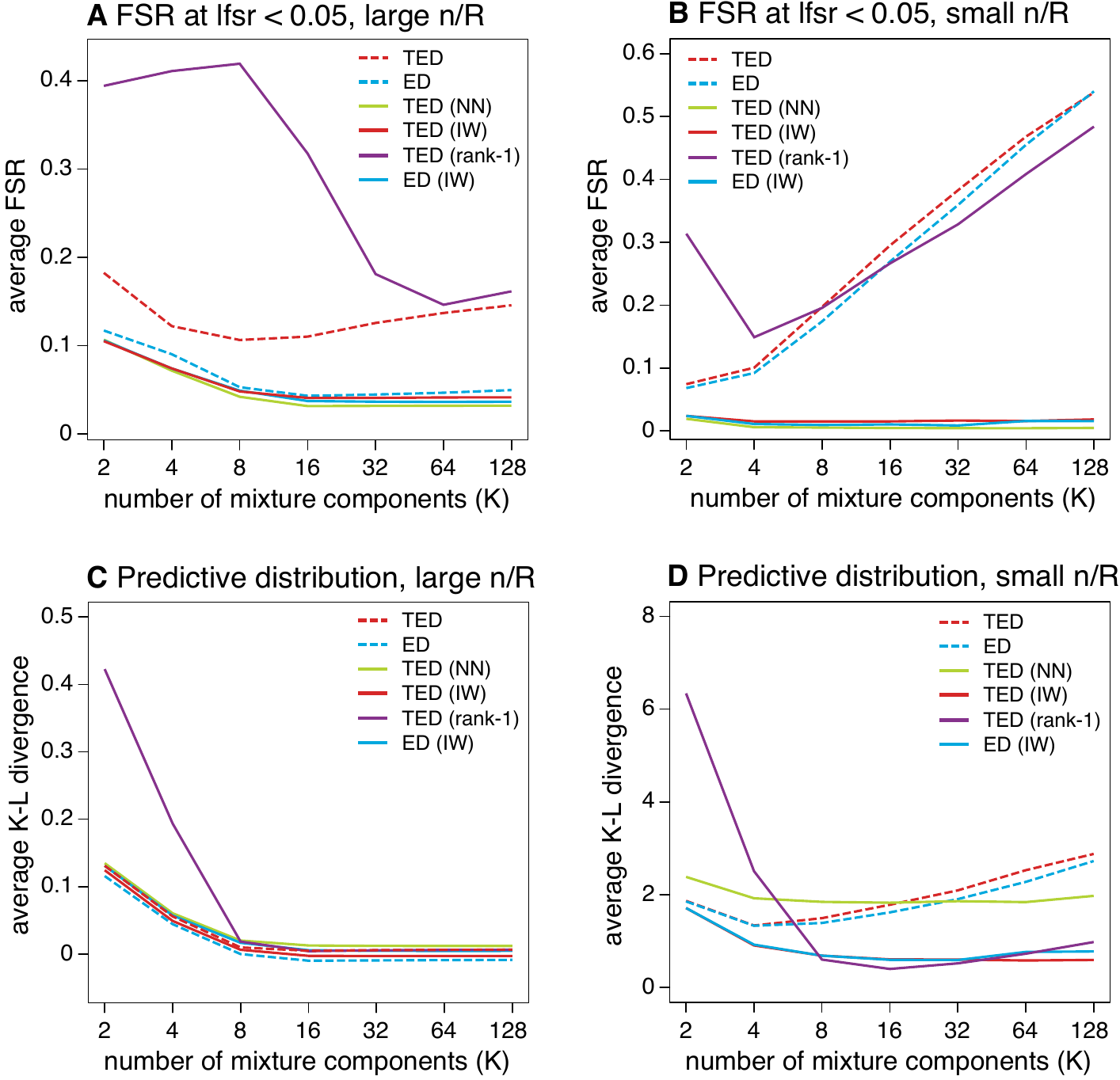}
\caption{\rm Assessment of robustness to mis-specifying $K$ in
  ``rank-1'' simulated data sets.
  All the results shown in the plots are averages over the 20 data
  sets. All data sets were simulated with $K = 10$ mixture
  components.}
\label{fig:robustK_2}
\end{figure}

In these experiments, we analyzed the same 80 data sets as in the
previous section (simulated with $K = 10$). We fit models with
different penalties, constraints and algorithms, with $K$ varying from
2 to 128.  We compared results in both the accuracy of the predictive
distribution (K-L divergence) and the average FSR at an {\em lfsr}
threshold of 0.05.

Results are shown in Figures \ref{fig:robustK} and
\ref{fig:robustK_2}. For large $n/R$, all model fits except those with
the rank-1 constraints were robust to overstating $K$, with similarly
good performance even at $K=128$. This is generally consistent with
the claim in \cite{urbut2019flexible} that the mixture prior should be
robust to overstating $K$.
However, for small $n/R$ the story is quite different: all of the
unpenalized algorithms eventually showed a decline in performance when
$K$ was too large, presumably due to ``overfitting''. In comparison,
all the penalized methods were more robust to overstating $K$; the
performance did not substantially decline as $K$ increased.
%
%

The improved robustness of the penalized methods could be achieved in
at least two different ways: they could be using a smaller number of
components by estimating some of the mixture weights $\pi_k$ to be
very small; or by estimating some of the components $k$ to have very
similar covariances $\U_k$ (or both). To investigate these
explanations, we looked at the models with $K = 100$ components and
recorded the number of ``important'' components, defined as components
$k$ with $\pi_k > 0.01$. We found that the penalized methods tended to
produce much fewer ``important'' components than the unpenalized
methods (Supplementary Figure \ref{supple:compare_weights}).
%
%
Essentially, the penalties have the effect of shrinking each $\U_k$
toward the identity matrix, so a component is assigned a small weight
whenever the identity matrix does not match the truth.

In summary, our results support the use of penalized methods with a
large value of $K$ as a simple and robust way to achieve good
performance in different settings.

\section{Analysis of genetic effects on gene expression in 49 human tissues}
\label{sec:application}

To illustrate our methods on real data, we used the EBMNM model to
analyze the effects of genetic variants on gene expression (``{\em
  cis-}eQTLs'') in multiple tissues. The use of the EBMNM model for
this purpose was first demonstrated in \cite{urbut2019flexible}. They
used a two-stage procedure to fit the EBMNM models: (i) fit the EBMNM
model to a subset of ``strong'' eQTLs to estimate the prior covariance
matrices; and (ii) fit a (modified) EBMNM model to all eQTLs---or a
random subset of eQTLs---using the covariance matrices from (i). The
model from (ii) was then used to perform inferences and to test for
eQTLs in each tissue. Our new methods are relevant to (i), and so we
focus on stage (i) here. For (i), \cite{urbut2019flexible} used the ED
algorithm without a penalty. Recognizing that the results are
sensitive to initialization, they described a detailed initialization
procedure.\footnote{An updated version of the initialization procedure
of \cite{urbut2019flexible} is described in the ``flash\_mash''
vignette included in the mashr R package. See also
\url{https://stephenslab.github.io/mashr/articles/flash_mash.html}.}
We coded an initialization procedure similar to this, which we refer
to as the ``specialized initialization''. We note that the specialized
initialization adds substantially to both the complexity and
computation time of the overall fitting procedure. Our goal was to
assess the benefits of different EBMNM analysis pipelines which
consisted of combinations of: TED vs. ED updates; specialized
initialization vs. a simple random initialization; and penalty
vs. no penalty (maximum-likelihood). All these combinations resulted
in 8 different analyses of multi-tissue cis-eQTL data.


We analyzed {\em z-}scores from tests for association between gene
expression in dozen human tissues and genotypes at thousands of
genetic variants. The {\em z-}scores came from running the ``Matrix
eQTL'' software \citep{matrix_eqtl} on genotype and gene expression
data in release 8 of the Genotype-Tissue Expression (GTEx) Project
\citep{gtex2020gtex}. Following \cite{urbut2019flexible}, we selected
the genetic variants with the largest {\em z-}score (in magnitude)
across tissues for each gene. After data filtering steps, we ended up
with a data set of {\em z-}scores for $n = \mbox{15,636}$ genes and $R
= 49$ tissues. We fit the EBMNM model to these data, with all the
$\V_j$ set to a common correlation matrix; that is, $\V_j = {\bm C}$,
where ${\bm C}$ is a correlation matrix of non-genetic effects on
expression. This correlation matrix ${\bm C}$ was estimated from the
association test {\em z-}scores following the approach described in
\cite{urbut2019flexible}. In all runs, we set $K = 40$ to match the
number of covariance matrices produced by our specialized
initialization. And, in all cases, we initialized the mixture weights
to $\pi_k = 1/40$ and the scaling factors (when needed) to
$s_k=1$. The supplementary ZIP file \citep{supplementary_zip} contains
the code and data implementing these analyses as well as the results
we generated.

\begin{figure}[t]
\centering
\includegraphics[width=\textwidth]{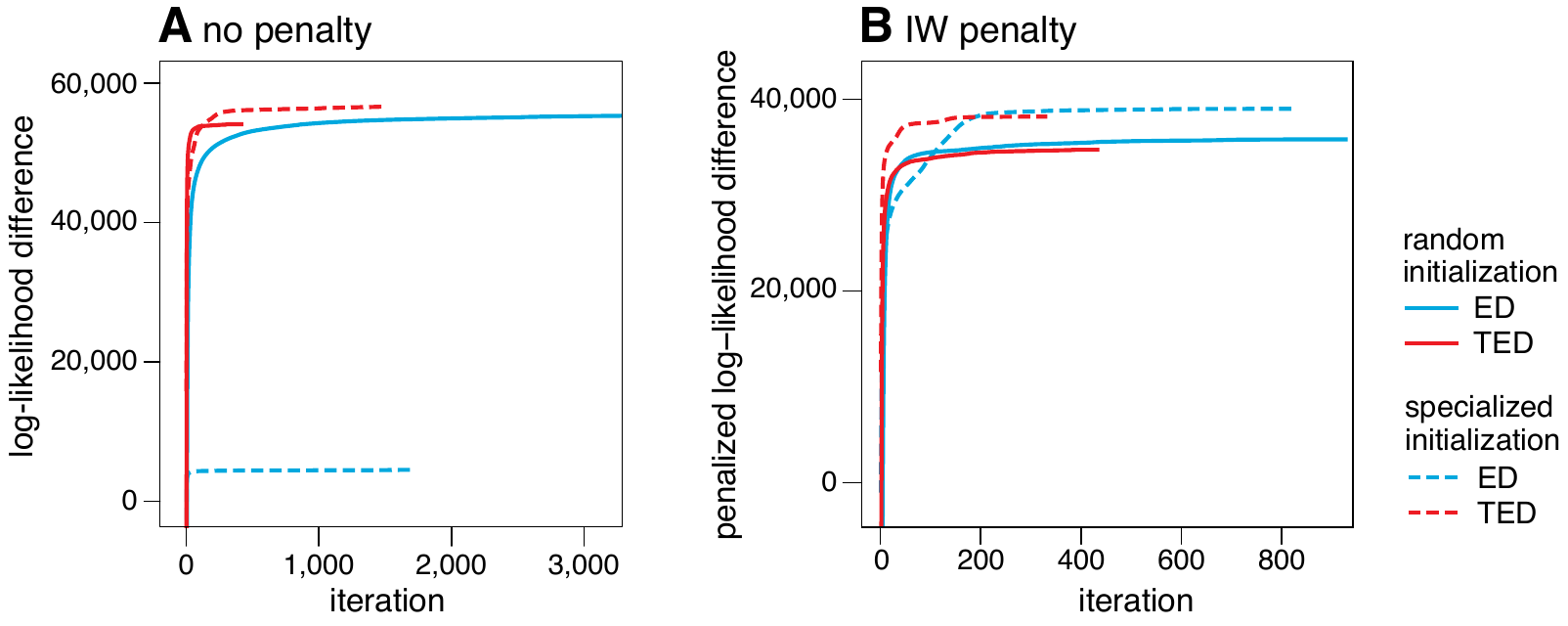}
\caption{\rm Plots showing improvement in model fit over time for the
  GTEx data, using different initialization schemes, different prior
  covariance matrix updates, and penalty vs. no penalty
  (maximum-likelihood). Log-likelihood differences and penalized
  log-likelihood differences are with respect to the (penalized)
  log-likelihood near the initial estimate. All models were fit with
  $K = 40$ mixture components. In B, the
  inverse wishart (IW) penalty was used with penalty parameter
  $\lambda = R$. In all cases, the model fitting was halted when the
  difference in the log-likelihood between two successive updates was
  less than 0.01, or when 5,000 updates were performed, whichever came
  first.}
\label{fig:fig1_real_dat}
\end{figure}

Figure \ref{fig:fig1_real_dat} shows the improvement of the model fits
over time in the 8 different analyses. In the analyses without a
penalty (A), the TED updates with a specialized initialization
produced the best fit, while the ED and TED updates with random
initialization were somewhat worse (e.g., TED with random
initialization produced a fit that was 2,497.78 log-likelihood units
worse, or 0.16 log-likelihood units per gene). Strikingly, the ED
updates with specialized initialization resulted in a much worse fit.
We attribute this to the fact that the specialized initialization
includes many rank-1 matrices, and the ``subspace preserving''
property of the ED updates means that these matrices are fixed at
their initialization (they changed only by a scaling factor), which
substantially limits their ability to adapt to the data.  In the
penalized case, consistent with our simulation results, there was less
difference between ED vs. TED. In both cases, the specialized
initialization improved the fit relative to random initialization
(e.g., TED increased the penalized log-likelihood by 3,176, or about 0.2 per gene). Note that adding a penalty function makes the subspace
preserving property of the ED updates irrelevant by forcing the
matrices to be full rank.

%
%
%


To compare the quality of the fits obtained by each method, we
computed log-likelihoods on held-out (``test set'') data, using a
5-fold cross-validation (CV) design. Following the usual CV setup, in
each CV fold 80\% of the genes were in the training set, and the
remaining 20\% were in the test set.
%
%
Then we fit an EBMNM model to the training set following each of the 8
approaches described above, and measured the quality of the fit by
computing the log-likelihood in the test set. We also recorded the
number of iterations. Within a given fold, the number of components
$K$ was the same across all the analyses, and was set depending on the
number of covariance matrices produced by the specialized
initialization. ($K$ was at least $32$ and at most $39$.)

Consistent with the simulations, the inclusion of a penalty
consistently improved the test set log-likelihood
(Table~\ref{table:transposed_test_data_loglik}). With a penalty, the
specialized initialization also improved the test set log-likelihood
compared with a random initialization. Of the 8 approaches tried, the
one used by \cite{urbut2019flexible}---ED with no penalty and
specialized initialization---resulted in the worst test-set
likelihood.

\begin{table}[t]
\centering
\caption{\rm Cross-validation results on the GTEx data.  The ``mean
  relative log-likelihood'' column gives the increase in the test-set
  log-likelihood over the worst log-likelihood among the 8 approaches
  compared, divided by total number of genes in each test set. The ``average
  number of iterations'' column gives the number of iterations
  performed until the stopping criterion is met (log-likelihood
  between two successive updates less than 0.01, up to a maximum of
  5,000 iterations), averaged over the 5 CV folds.}
\label{table:transposed_test_data_loglik}
\begin{tabular}{@{}c@{\;\;}c@{\;\;}c@{\;\;}r@{\;\;}r@{}}
\toprule
& & & mean relative & average number \\[-0.3ex]
initialization & algorithm & penalty & log-likelihood &
of iterations \\
\midrule
specialized & ED & none & 0.00 & 1,101 \\
specialized & ED & IW & 1.21 & 1,083 \\
specialized & TED & none & 0.88 & 1,054 \\
specialized & TED & IW & 1.19 & 412 \\
random & ED & none & 0.25 & 5,000 \\
random & ED & IW & 0.86 & 1,377 \\
random & TED & none & 0.20 & 450 \\
random & TED & IW & 0.94 & 584 \\
\bottomrule
\end{tabular}
\end{table}

\begin{figure}[t]
\centering
\includegraphics[width=0.45\textwidth]{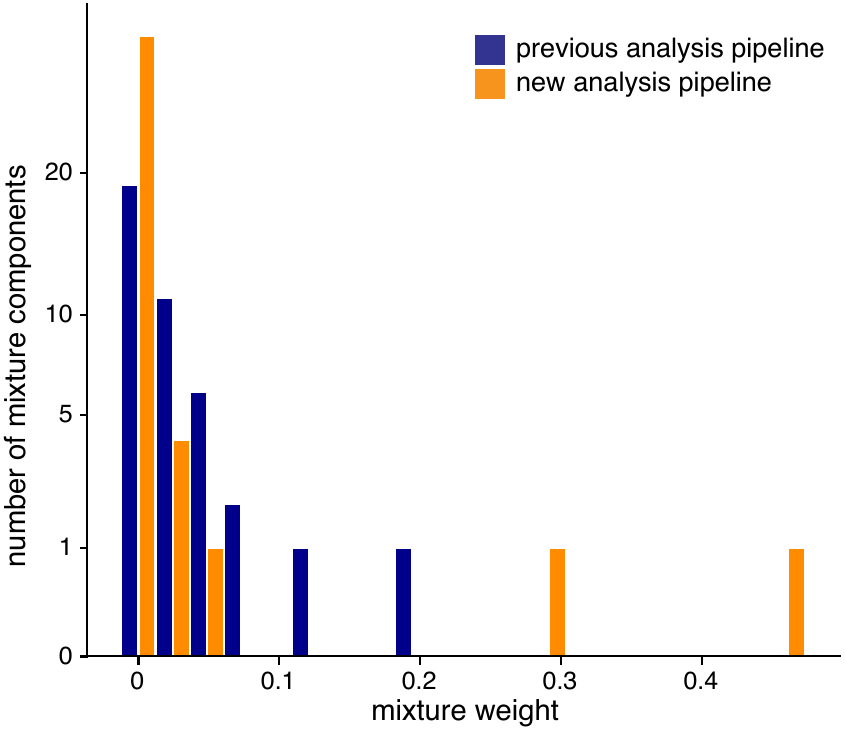}
\caption{\rm Comparison of the prior mixture weights $\bpi$ from the
  previous pipeline vs. the new pipeline. The histogram shows the
  distributions of the $K 40$ prior mixture weights resulting from both
  pipelines.}
\label{fig:gtex_mixture_weights}
\end{figure}

\begin{figure}[t]
\centering
\includegraphics[width=\textwidth]{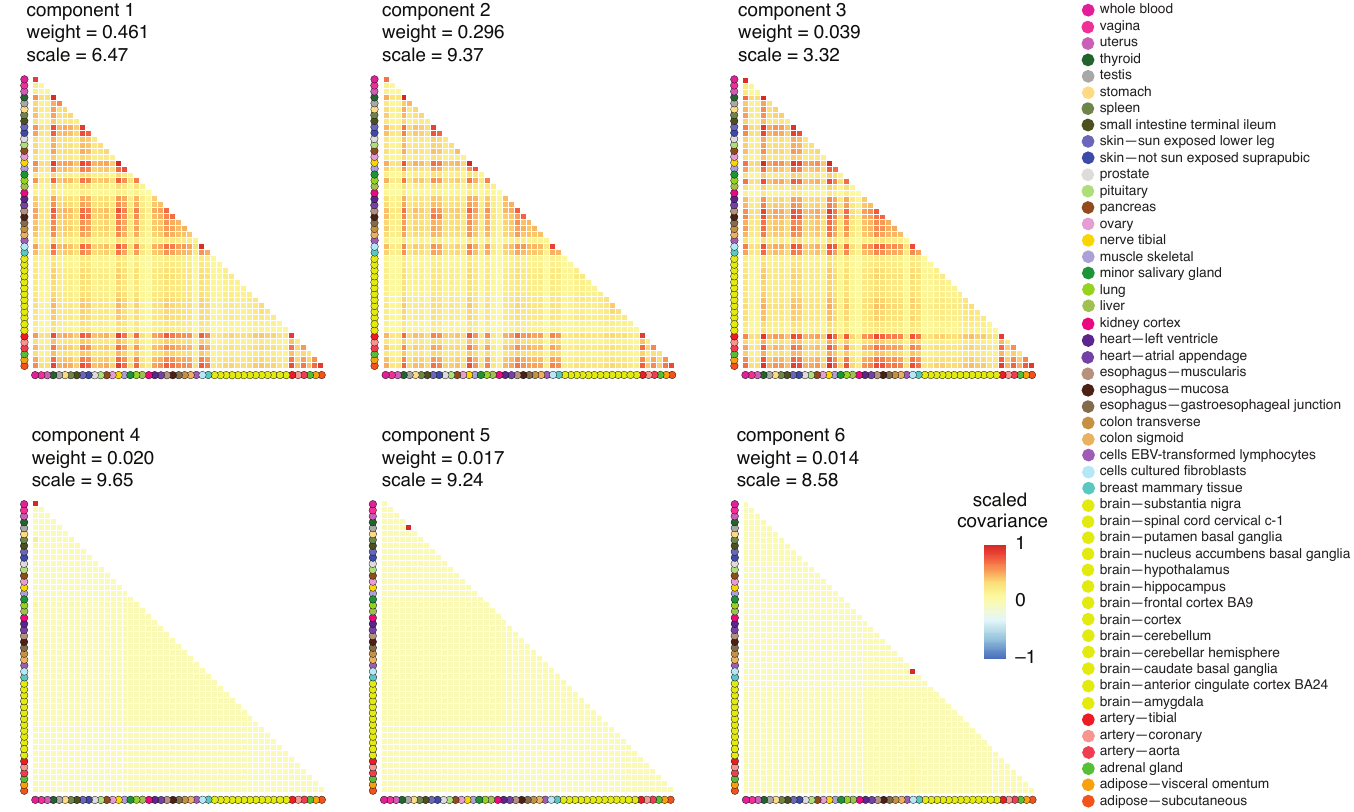}
\caption{\rm Top effect-sharing patterns in the EBMNM model fit to the
  GTEx data using the previous pipeline. The ``top'' patterns are the
  mixture components with the largest weights. Each heatmap shows the
  49 $\times$ 49 the scaled covariance matrix $\U_k/\sigma_k^2$, where
  $\sigma_k^2$ is the largest diagonal element of $\U_k$, so that all
  elements of the scaled covariance lie between $-1$ and 1. (The vast
  majority of the covariances are positive; negative correlations are
  unexpected.) The scaled covariance matrices are arranged in
  decreasing order by mixture weight. The ``scale'' above each heatmap
  is $\sigma_k$. Note that the top three covariance matrices capture
  broadly similar effect-sharing patterns, but different effect
  scales.}
\label{fig:gtex_covariances}
\end{figure}

\begin{figure}[t]
\centering
\includegraphics[width=0.965\textwidth]{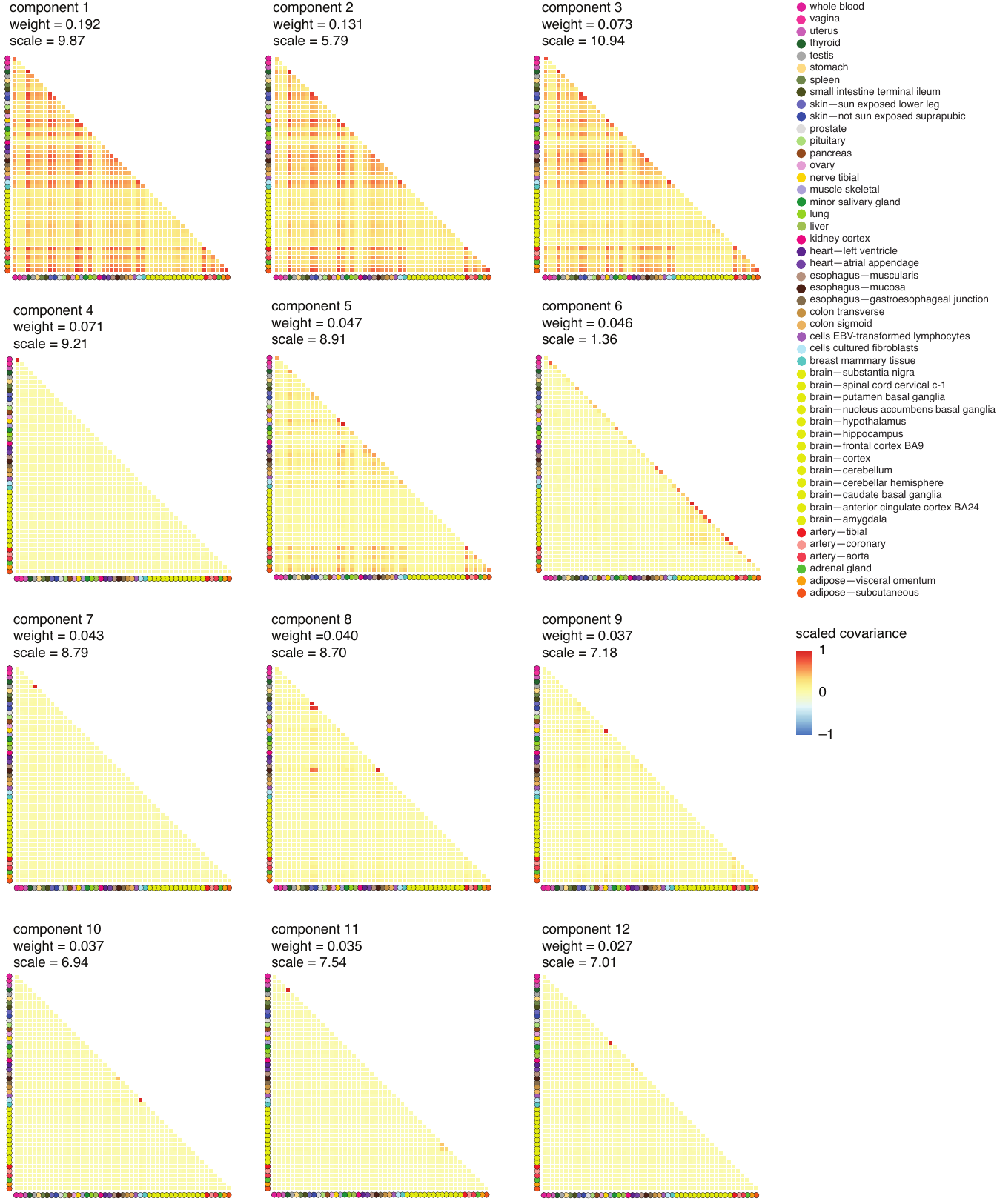}
\caption{\rm Top effect-sharing patterns in the GTEx data generated by
  the new analysis pipeline. See the caption to
  Fig.~\ref{fig:gtex_covariances} for more details.}
\label{fig:gtex_covariances_new}
\end{figure}

Although the analyses with a specialized initialization resulted in
better fits than the analyses with a random initialization, the
specialized initialization also had substantial computational
overhead; running the initialization procedures on these data took
more than 1 hour (by comparison, running a single iteration of the
EBMNM algorithm typically took about 1 second). Therefore, on balance,
one might prefer to dispense with the specialized
initialization. Based on these results and considerations, we
subsequently examined in more detail the analysis with ED, no penalty
and specialized initialization---which was the approach used in
\cite{urbut2019flexible}---and the analysis with TED, IW penalty and
random initialization. For brevity, we refer to these analyses as the
``previous pipeline'' and ``new pipeline'', respectively.

A notable outcome of the new pipeline is that it produced a prior with
weights that were more evenly distributed across the mixture
components (Figure~\ref{fig:gtex_mixture_weights}). For example, the
new pipeline produced 22 covariances with weights greater than 1\%,
whereas the previous pipeline produced only 10 covariances with
weights more than 1\%. Also, the top 16 covariances accounted for 85\%
of the total weight in the new pipeline, whereas only 6 covariances
were needed to equal 85\% total weight in the previous
pipeline. Inspecting the individual covariances generated from the two
analysis pipelines, there many strong similarities in the estimated
sharing patterns (Figures \ref{fig:gtex_covariances} and
\ref{fig:gtex_covariances_new}). But the new analysis pipeline learned
a greater variety of tissue-specific patterns (e.g., whole blood,
testis, thyroid) and tissue-sharing patterns, many of which appear to
reflect underlying tissue biology. For example, sharing pattern 6 (see
Figure~\ref{fig:gtex_covariances_new}) captures sharing of
brain-specific effects (including the pituitary gland, which is found
at the base of the brain near the hypothalamus). Sharing pattern 8 may
reflect the fact that skin and the mucosa layer of the esophagus wall
both contain squamous epithelial cells. Additionally, 
we looked at more closely at the 8 genes
with very strong posterior weights (>98\%) 
on component 6, the component capturing eQTL sharing among brain tissues:
{\em HOMER1}, {\em JPT1},
{\em SRSF2}, {\em ABCA1}, {\em GPATCH8}, {\em SYNGAP1}, {\em SPI1} and
{\em LGMN.}
Several of these 
have been linked to neurological and neuropsychiatric conditions, 
including major depression, schizophrenia, attention
dementia and Alzheimer's disease. 
For example, {\em SYNGAP1} is linked to
neuronal functions and psychiatric diseases based on results 
from the GWAS
Catalog \citep{gwas_catalog} and from functional studies
\citep{jeyabalan2016syngap1, llamosas2020syngap1}. In summary, the
improvements to the EBMNM analysis pipeline should result in the
discovery of a greater variety of cross-tissue genetic effects on gene
expression.

\section{Discussion}
\label{sec:discussion}

The growing interest in deciphering shared underlying biological
mechanisms has led to a surge in multivariate analyses in genomics;
among the many recent examples are multi-trait analyses
\citep{wu2020multi, luo2020multi} and multi-ancestry analyses to
improve polygenic risk scores \citep{zhang2023new}. The EBMNM approach
described here and in \cite{urbut2019flexible} provide a versatile and
robust multivariate approach to multivariate analysis. We have
implemented and compared several algorithms for this problem. These
algorithms not only enhance accuracy but also provide a level of
flexibility not achieved by other methods.

One of our important findings is that using low-rank covariance
matrices in this setting, as was done in \cite{urbut2019flexible}, is
not recommended. In particular, while low-rank matrices may be
relatively easy to interpret, they lead to poorly calibrated {\em
  lfsr} values (e.g., Figure \ref{fig:inference_rank1}). For
intuition, consider fitting a EBMNM model with $K = 1$ and a rank-1
covariance matrix, $\U_1 = \bs{u}\bs{u}^T$. (We credit Dongyue Xie for
this example.) Under this model, the mean is $\btheta_j = \bs{u} a_j$
for some $a_j$, and therefore, given a $\bs{u}$, the signs of all the
elements of $\btheta_j$ are fully determined by $a_j$. As a result,
all elements of $\btheta_j$ will have the same {\em lfsr}, and so this
model cannot capture situations where one is confident in the sign of
some elements of $\btheta_j$ but not others.
This can cause problems {\em even if the model is correct}, that is,
when the true covariances are rank-1, as in Figure
\ref{fig:inference_rank1}. There are some possible ways to address
these issues, say, by imposing sparsity on estimates of $\bs{u}$, and
this could be an area for future work.

Even when the pitfalls of low-rank matrices are avoided, it is still
the case that EB methods tend to understate uncertainty compared to
``fully Bayesian'' methods (e.g., \citealt{morris1983parametric,
  wang2005inadequacy}). As a result, one should expect that the estimated
{\em lfsr} values may be anti-conservative; that is, the {\em lfsr}
values are smaller than they should be. Indeed, we saw this
anti-conservative behavior across many of our simulations and methods
(Supplementary Figures \ref{supple:calibration} and
\ref{supple:calibration_rank1}). For this reason estimated {\em lfsr}
rates should be treated with caution, and it would be prudent to use
more stringent significance thresholds than are actually desired (e.g.,
an {\em lfsr} threshold of 0.01 rather than 0.05). In the special
case where $\V=\I$, it should be possible to improve calibration of
significance tests by using ideas from \cite{lei2018adapt}. Improving
calibration in the general case of dependent multivariate tests seems
to be an important area for future research.

\begin{acks}[Acknowledgments]
We thank Gao Wang and Sarah Kim-Hellmuth for their help in preparing
the GTEx data. We thank Yuxin Zou, Yusha Liu and Dongyue Xie for
helpful discussions on MASH and the mashr software. And we thank the
staff at the Research Computing Center at the University of Chicago
for maintaining the computing resources used in our numerical
experiments.
\end{acks}

\begin{funding}
This work was supported by NIH grant R01HG002585 to MS.
\end{funding}

\begin{supplement}

\stitle{Supplementary materials}

\sdescription{The supplementary PDF includes supporting mathematical
  results and additional tables and figures.}

\end{supplement}

\begin{supplement}
  
\stitle{R package and code reproducing the analyses}
  
\sdescription{ZIP file containing the source code for the R package as
  well as the source code reproducing the results of the simulations
  and analysis of the GTEx data.}
  
\end{supplement}

\bibliographystyle{imsart-nameyear}
\bibliography{udr}

\end{document}